\documentclass[12pt]{article}
\pdfoutput=1
\usepackage{amsmath,amssymb,amsfonts}
\usepackage{a4wide,epsfig,psfrag,scalefnt}
\usepackage[dvipsnames]{xcolor}
\usepackage{braket}
\usepackage{placeins}
\usepackage{breqn}
\usepackage{slashed}
\usepackage{enumitem}
\usepackage[numbers,sort&compress]{natbib}
\usepackage{caption}
\usepackage{subcaption}
\usepackage{dsfont}%use \mathds{1} to get identity symbol
\usepackage{url}
\usepackage{hyperref}
\usepackage{booktabs}

\graphicspath{ {figs/} }

\allowdisplaybreaks

\begin{document}

\title{\vskip-3cm{\baselineskip14pt
    \begin{flushleft}
     \normalsize P3H-24-031, TTP24-013, SI-HEP-2024-11
    \end{flushleft}} \vskip1.5cm
  $B$ meson mixing at NNLO: technical aspects
}

\author{
  Pascal Reeck$^{a}$,
  Vladyslav Shtabovenko$^{b}$,
  Matthias Steinhauser$^{a}$,
  \\
  {\small\it (a) Institut f{\"u}r Theoretische Teilchenphysik,
    Karlsruhe Institute of Technology (KIT),}\\
  {\small\it Wolfgang-Gaede Strasse 1, 76131 Karlsruhe, Germany}
  \\
  {\small\it (b) Theoretische Physik 1, Center for Particle Physics Siegen,}\\ 
  {\small\it Universit{\"a}t Siegen, Walter-Flex-Str. 3, 57068 Siegen, Germany}
}

\date{}

\maketitle

\thispagestyle{empty}

\begin{abstract}

  We provide details to several technical aspects which are important
  for the calculation of next-to-next-to-leading order corrections
  to the mixing of neutral $B$ mesons. This includes the
  computation of the master integrals for finite charm and bottom quark
  masses, traces over products of up to 22 $\gamma$ matrices and
  tensor integrals with up to rank 11.
  
\end{abstract}

%- }}}

\newpage

%- {{{ Introduction and notation:

\section{Introduction and notation}
In the Standard Model of particle physics, the
mixing of neutral $B$ mesons is a loop-induced process which, at leading-order, is mediated by two box Feynman diagram where quark lines exchange two virtual $W$ bosons.
The time evolution is governed by a $2\times 2$ matrix
which contains the mass and decay matrices $M^q$ and $\Gamma^q$
with $q=s,d$ being the flavour of the spectator quark.

The quantities which are measured by experiment are the
mass and width differences of the lighter and heavier eigenstates,
\begin{eqnarray}
   \Delta M_q &=& M_H^q - M_L^q\,,\nonumber\\
   \Delta \Gamma_q &=& \Gamma_L^q - \Gamma_H^q\,,\label{eq:dg}
\end{eqnarray}
and the charge-parity (CP) asymmetry in flavour-specific decays,
$a_{\rm fs}^q$.  They are related to the matrix elements of the decay
matrix via (see, e.g., Ref.~\cite{Lenz:2006hd} for more details)
\begin{eqnarray}
  \Delta M_q  &=& 2|M_{12}^q| + \mathcal{O}(|\Gamma_{12}|^2/|M_{12}|^2)\,,\nonumber\\
  \frac{\Delta\Gamma_q}{\Delta M_q } &=&
                 - \mbox{Re}\frac{\Gamma_{12}^q}{M_{12}^q}\,,
                 \nonumber\\
  a_{\rm fs}^q &=& \mbox{Im} \frac{\Gamma_{12}^q}{M_{12}^q}\,,
                 \label{eq:dgdmafsq}
\end{eqnarray}
where $M_{12}^q$ and $\Gamma_{12}^q$ are obtained from the dispersive and absorptive parts of the $B_q\to \overline{B}_q$ transition amplitude, respectively.
In this work we concentrate on the computation of $\Gamma_{12}^q$.

The theoretical framework for the computation of $\Gamma^q_{12}$ is given by
a tower of effective theories. Using the Standard Model as starting
point, one first integrates out the top quark mass, Higgs boson and
electroweak gauge bosons and obtains the so-called $|\Delta B|=1$
effective theory (cf. e.g. Ref.~\cite{Buras:2020xsm} for more details). Within this theory one considers the correlator
of two $\Delta B=1$ operators which mediates the $B- \overline{B}$
transition.\footnote{See Appendix~\ref{app::ops} for a list of all relevant operators.}  Subsequently, one performs a heavy mass expansion
which leads to (see again Ref.~\cite{Lenz:2006hd} for a detailed derivation)
\begin{eqnarray}
	\Gamma_{12}^q &=& - (\lambda_c^q)^2\Gamma^{cc}_{12} 
	- 2\lambda_c^q\lambda_u^q \Gamma_{12}^{uc} 
	- (\lambda_u^q)^2\Gamma^{uu}_{12} 
                          \nonumber\\
                      &=& -(\lambda_t^q)^2 \left[
                          \Gamma_{12}^{cc} 
                          + 2 \frac{\lambda_u^q}{\lambda_t^q}\left(\Gamma_{12}^{cc}-\Gamma_{12}^{uc}\right)
                          + \left(\frac{\lambda_u^q}{\lambda_t^q}\right)^2 
                          \left(\Gamma_{12}^{uu}+\Gamma_{12}^{cc}-2\Gamma_{12}^{uc}\right)
                          \right]
	\,,
	\label{eq::Gam12}
\end{eqnarray}
where $\lambda^q_a = V_{aq}^\ast V_{ab}$ and $\Gamma_{12}^{ab}$ follow
from the absorptive part of a bi-local matrix element containing a
time-ordered product of two $|\Delta B| = 1$ effective
Hamiltonians. This yields
\begin{eqnarray}
	\Gamma_{12}^{ab} 
	&=& \frac{G_F^2m_b^2}{24\pi M_{B_s}} \left[ 
	H^{ab}(z)   \langle B_s|Q|\bar{B}_s \rangle
	+ \widetilde{H}^{ab}_S(z)  \langle B_q|\widetilde{Q}_S|\bar{B}_q \rangle
	\right]
	+ \mathcal{O}(\Lambda_{\rm QCD}/m_b) \,,
	\label{eq::Gam^ab}
\end{eqnarray}
where terms suppressed by the inverse bottom quark mass are
not shown. The precise definition of the operators $Q$ and $\widetilde{Q}_S$ can be found in Appendix~\ref{app::ops}. $\langle B_s|Q|\bar{B}_s \rangle$ and $\langle
B_q|\widetilde{Q}_S|\bar{B}_q \rangle$ are non-perturbative matrix
elements, which are computed using lattice gauge
theory~\cite{Dowdall:2019bea} or QCD sum
rules~\cite{Kirk:2017juj,King:2021jsq}.  The perturbative matching
coefficients $H^{ab}(z)$ and $\widetilde{H}^{ab}_S(z)$ are the main focus of this paper. They depend on the
mass ratio
\begin{eqnarray}
 z &=& \frac{m_c^2}{m_b^2}\,.
\end{eqnarray}
NLO QCD corrections to the matching coefficients with current-current
operators have been known for quite some
time~\cite{Beneke:1998sy,Ciuchini:2003ww,Beneke:2003az,Lenz:2006hd}.
The complete set of penguin operators has only been added
recently~\cite{Gerlach:2021xtb,Gerlach:2022wgb}, where terms of order $z^0$ and $z^1$ have been computed in an expansion for $z\to0$. Fermionic NNLO
corrections have been considered in
Refs.~\cite{Asatrian:2017qaz,Asatrian:2020zxa,Hovhannisyan:2022miy}
and the complete NNLO contribution from current-current operators is
known from Ref.\cite{Gerlach:2022hoj}, however, only in an expansion
for small charm quark masses up to (including) ${\cal O}(z)$.  The
current experimental and theoretical uncertainties (see
Ref.~\cite{Gerlach:2022hoj} for a recent compilation) are such, that a
complete NNLO calculation involving all penguin contributions for
$H^{ab}(z)$ and $\widetilde{H}^{ab}_S(z)$ is necessary.
There exist several challenges that make this endeavour non-trivial:
\begin{enumerate}
\item Large number of contributing Feynman diagrams.
\item Traces involving a large number of $\gamma$ matrices.
\item High-rank tensor integrals.
\item Integration-by-parts reduction to master integrals.
\item Computation of the master integrals.
\item Proper inclusion of evanescent operators.
\end{enumerate} 

In this paper we address the points 2, 3 and 5.  The large number of
diagrams is not a fundamental issue, since it is always possible to split the
problem into smaller pieces during the generation of the Feynman amplitudes
with {\tt qgraf}~\cite{Nogueira:1991ex}. For the integration-by-parts reduction we use
{\tt Kira}~\cite{Maierhofer:2017gsa,Maierhofer:2018gpa,Klappert:2020nbg} which requires at most a few days for general gauge parameter. Thus, this is not a limiting factor of our calculation.  The last point concerning the evanescent operators is a purely field-theoretical problem. It will be addressed in detail in Ref.~\cite{Nierste_etal_inprep}.

The remainder of the paper is organized as follows: In the next Section we describe in detail the construction of efficient projectors and Section~\ref{sec:tensor} deals with the complementary approach based on high-rank tensor integrals.
In Section~\ref{sec:master} we discuss our methods to compute the 
master integrals. A sample result where the methods of Sections~\ref{sec:projectors},~\ref{sec:tensor} and~\ref{sec:master} have been applied is shown in Section~\ref{sec:results}. We conclude in Section~\ref{sec:concl}. In Appendix~\ref{app:basis} the complete list of basis elements relevant for our calculation is shown and Appendix~\ref{app::ops} contains a complete list of relevant operators of the $\Delta B=1$ and $\Delta B=2$ theory.

%- }}}
%- {{{ Long traces:

\section{Projector methodology}
\label{sec:projectors}

\subsection{Problem statement}
\begin{figure}[t]
    \centering
    \includegraphics[scale=0.6]{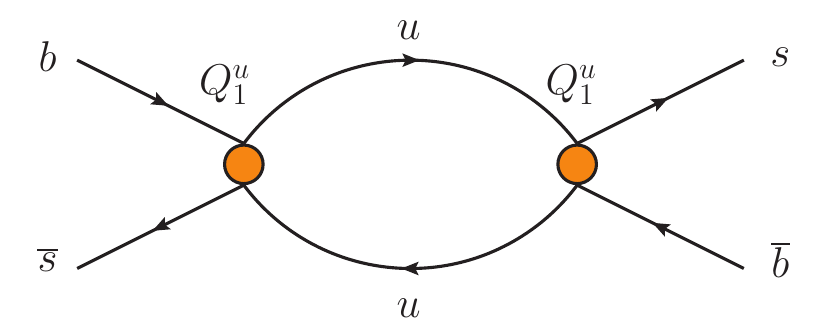}
    \caption{A simple one-loop example for the operator combination $Q_1^{u}\times Q_1^{u}$.}
    \label{fig:simple_1l}
\end{figure}

We wish to apply a set of projectors to our amplitude in order to decompose the tensor integrals into products of scalar integrals, colour structures and spinor structures which are free of loop momenta. The colour part of our diagrams can be treated separately and is straightforward to handle with projectors as there are only two tensor structures corresponding to the two possible colour contractions of the external quarks. The spinor structures encountered in diagrams before applying projectors consist of two matrices in Dirac spinor space connecting a pair of external quarks each. While these objects have a simple structure in spinor space, they possess a more complicated tensor structure in Lorentz space since they are composed of a product of Dirac $\gamma$ matrices, which we will call \textit{spin lines} or \textit{Dirac chains}. The Lorentz indices are contracted either across the two spin lines or with a propagator momentum. For example, the term
\begin{equation}
    P_R \,\gamma^\mu\, \slashed{p_1}\, \gamma^\nu \otimes P_R \,\gamma_\nu\, \slashed{p_2}\, \gamma_\mu
\end{equation}
corresponds to the spinor part of the diagram shown in Fig.~\ref{fig:simple_1l}. Here, $p_1$ and $p_2$ are linear combinations of loop and external momenta.
Note that the chirality projector $P_R=(1+\gamma_5)/2$ can always be commuted to the end of a spin line since we are using the basis of Ref.~\cite{Chetyrkin:1997gb}. Thus, in our calculation, $\gamma_5$ never appears in closed fermion loops and we can use anti-commuting $\gamma_5$.

\begin{figure}
    \centering
    \begin{tabular}{ccc}
    \includegraphics[scale=0.35]{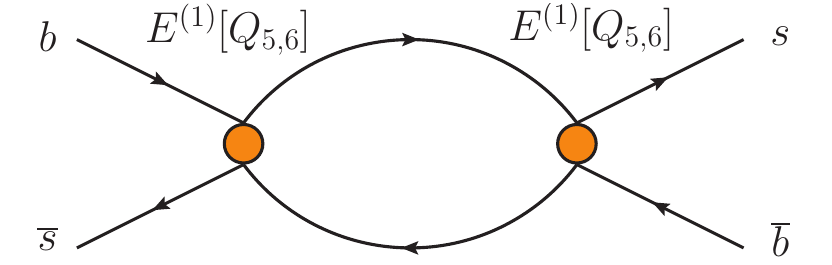}&
    \includegraphics[scale=0.35]{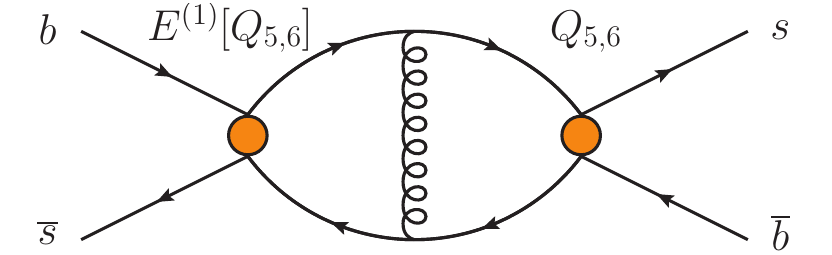}&
    \includegraphics[scale=0.35]{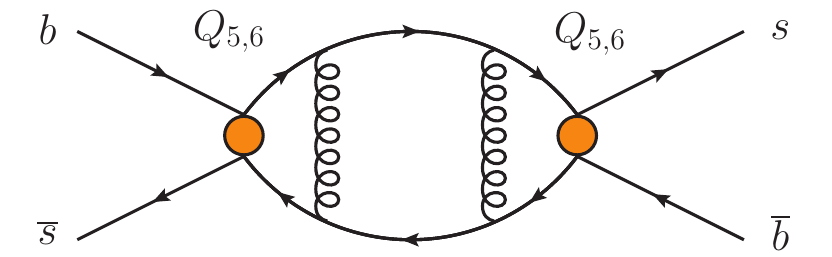}\\
    (a) & (b) & (c)
    \end{tabular}
    \caption{Sample diagrams at one-loop (a), two-loop (b) and three-loop (c) order which will give rise to the largest number of $\gamma$ matrices in our amplitude: eleven.}
    \label{fig:sample_diagrams}
\end{figure}

The maximum number of $\gamma$ matrices in our diagrams is eleven on each of the two spin lines, which includes at least one slashed momentum per spin line. Such terms appear, e.g., in the following diagrams:
\begin{itemize}
\item To one-loop order, we have diagrams with two evanescent operators of the first generation which have up to five $\gamma$ matrices for the penguin operators, see Fig.~\ref{fig:sample_diagrams}(a). The amplitudes with the most $\gamma$ matrices have ten $\gamma$ matrices with open Lorentz indices as well as one slashed momentum per spin line. We will call $\gamma$ matrices which do not have their Lorentz index contracted with a propagator momentum \textit{pure}.
\item To two-loop order, we have diagrams with one first generation evanescent operator and one penguin operator with an additional gluon. The penguin operators have up to three $\gamma$ matrices while the evanescent operator yields at most five $\gamma$ matrices. The longest chain of $\gamma$ matrices consists of nine pure $\gamma$ matrices and two slashed momenta per spin line with a gluon connecting the two spin lines, see Fig.~\ref{fig:sample_diagrams}(b).
\item To three-loop order, we have diagrams with two penguin insertions and two gluons. The longest spin lines result from gluons connecting across the two as shown in Fig.~\ref{fig:sample_diagrams}(c), so we have at most eight pure $\gamma$ matrices and three slashed momenta per spin line.
\end{itemize}

We will proceed with some details on the underlying mathematics, describing the occurring structures and defining a proper set of projectors. An efficient algorithm for loop calculations based on those considerations can be found in Section \ref{sec:algorithm}.

\subsection{Vector spaces spanned by \texorpdfstring{\boldmath$\gamma$}{gamma} matrices}

\subsubsection{Vector spaces from pure \texorpdfstring{\boldmath$\gamma$}{gamma} matrices}
\label{sec:pure_vector_spaces}

Focusing on pure $\gamma$ matrices first, we encounter spinor structures in the amplitude which consist of the tensor product of two spin lines,
\begin{equation}
    \Xi = \Gamma^{\mu_1,\dots,\mu_n} \otimes \Gamma_{\mu_{\sigma(1)},\dots,\mu_{\sigma(n)}},
\end{equation}
where $\sigma(x)$ is a particular permutation of the $n$ indices. Any index contractions on the same spin line have been taken care of already by commuting the $\gamma$ matrices together. We can commute the $\gamma$ matrices in accordance with the Clifford algebra,
\begin{equation}
    \{\gamma^\mu, \gamma^\nu\} = 2 g^{\mu\nu},
    \label{eq:clifford_algebra}
\end{equation}
where we have chosen the naive dimensional regularisation scheme. This means that we simply treat metric tensor to be in $d$ dimensions and work with an anticommuting $\gamma_5$. Although the spinor index on the $\gamma$ matrices becomes formally infinite-dimensional, the trace of the unit matrix can still be chosen to be equal to four. At the fundamental level, the spinor structures are elements of
\begin{equation}
    \mathit{Cl}_{p,q}(\mathbb{C}) \otimes \mathit{Cl}_{p,q}(\mathbb{C}),
\end{equation}
where the spacetime is $\mathbb{R}^{p,q}$ with $p=1$ time and $q=d-1$ space dimensions. Since the Clifford algebra is not closed in $d$ dimensions, we would like to further narrow down the structures that we consider and the vector space that they live in. In the following we will take a bottom-up approach by only considering the spinor structures that appear in our amplitude.

A single diagram generates Dirac chains of different lengths, and we note that the elements with different lengths of $\gamma$ matrices are components of an $n$-tuple
\begin{equation}
    x = \left(c^{(0)} \mathds{1}\otimes\mathds{1},\, c^{(1)} \gamma^\mu\otimes\gamma_\mu, \,\sum_i c^{(2)}_i \Gamma^{(2)}_i \otimes \Gamma^{(2)}_i,\, \dots\right),
    \label{eq:n_tuple_gammas}
\end{equation}
where $\Gamma^{(k)}_i$ refers to a string of $k$ $\gamma$ matrices for a particular permutation of the indices. Note that the first spin line can always be canonically labelled with indices $\mu_1, \dots, \mu_n$; the permutation refers to indices on the second spin line. Each diagram then involves one such $n$-tuple which lives in the vector space $\tilde{V}_n$  formed by the direct sum of the $k$-dimensional vector spaces $\tilde{V}^{(k)}$ of unordered $\gamma$ matrices of length $k$, up to a maximum length $n$,
\begin{equation}
    \tilde{V}_n \equiv \tilde{V}^{(0)} \oplus \tilde{V}^{(1)} \oplus \, \dots \, \oplus \tilde{V}^{(n)}.
\end{equation}
However, there is some redundancy in this description as we can use the Clifford algebra from Eq.~\eqref{eq:clifford_algebra} to commute the $\gamma$ matrices into a canonical order. This provides a linear map $\phi_{\text{Cliff}}: \tilde{V}_n \to V_n$, where $V_n$ is the vector space of ordered spin lines
\begin{equation}
    V_n = \text{span}_{\mathbb{C}} \left( \{\mathds{1} \otimes \mathds{1}, \gamma^\mu \otimes \gamma_\mu,\, \dots\, , \gamma^{\mu_1} \dots \gamma^{\mu_n} \otimes  \gamma_{\mu_n} \dots \gamma_{\mu_1} \} \right).
    \label{eq:gamma_basis}
\end{equation}
Here, we have chosen one particular ordering of the $\gamma$ matrices for every possible length of the spin lines and spanned a vector space by allowing each basis element to have arbitrary complex coefficients. The elements above are linearly independent in $d$ dimensions because we cannot use four-dimensional identities of the Dirac algebra like the Chisholm identities to reduce the number of $\gamma$ matrices on a spin line. Note that we can rewrite our vector space $V_n$ as the direct sum of the one-dimensional vector spaces $V^{(k)}$ of ordered $\gamma$ matrices of length $k$, up to a maximum length $n$,
\begin{equation}
    V_n \equiv V^{(0)} \oplus V^{(1)} \oplus \, \dots \, \oplus V^{(n)}, \label{eq:vector_space_ordered}
\end{equation}
which is isomorphic to $\mathbb{C}^{n+1}$ due to the isomorphism of the vector spaces $V^{(k)} \cong \mathbb{C}$. The set of vectors spanning $V_n$ as defined in Eq.~\eqref{eq:gamma_basis} constitutes a basis since they are linearly independent. Furthermore, we have $\tilde{V}^{(r)} \subset \tilde{V}^{(s)} \,\forall \, r < s$ and $V_n \subset \tilde{V}_n$. For Dirac chains of pure $\gamma$ matrices, the task we are faced with, is to resolve the map $\phi_{\text{Cliff}}$ in an efficient manner.

\subsubsection{Including slashed momenta}

As a reordering of the $\gamma$ matrices in a spin line is always possible using Eq.~\eqref{eq:clifford_algebra}, we will consider ordered structures of the form
\begin{equation}
    \Xi = \slashed{p}_1\dots\slashed{p}_m\,\gamma^{\mu_1} \dots \gamma^{\mu_n} \otimes \slashed{p}_{m+1}\dots\slashed{p}_k\,\gamma_{\mu_n} \dots \gamma_{\mu_1},
    \label{eq:slashed_sample_element}
\end{equation}
where we have assumed that the map between $\tilde{V}_n$ and $V_n$ described in Section \ref{sec:pure_vector_spaces} has been resolved and the momenta have been commuted to one end of the spin lines already. As will be explained in Section \ref{sec:algorithm}, this will turn out to be an efficient way of tackling the problem. Note that the number of momenta $k$ that can maximally occur has an upper bound from the number of propagators in the diagram.

Formally, the structures in Eq.~\eqref{eq:slashed_sample_element} are elements of the vector space
\begin{equation}
    \Xi \in \mathbb{C}^d \oplus \dots \oplus \mathbb{C}^d \oplus V_n \cong \left(\mathbb{C}^d\right)^k \oplus V_n \cong \left(\mathbb{C}^d\right)^k \oplus \mathbb{C}^{n+1} \equiv U_{n,k}',
\end{equation}
where the dimension of the Lorentz index is $d=4-2\epsilon$ and $\cong$ denotes an isomorphism. However, we know that the final result of carrying out the loop integral can only give a limited number of tensor structures in Lorentz space because the only available tensors are $q^\mu$, the external momentum, and $g^{\mu\nu}$, the metric tensor. Therefore, we are interested in finding a projector which maps onto the smaller vector space
\begin{equation}
    U_{n,k} \equiv \text{span}_{\mathbb{C}} \left( \{\mathds{1} \otimes \mathds{1}, \slashed{e}_q \otimes \mathds{1}, \mathds{1} \otimes \slashed{e}_q, \slashed{e}_q \otimes \slashed{e}_q, \gamma^\mu \otimes \gamma_\mu,\, \dots\, , \gamma^{\mu_1} \dots  \gamma^{\mu_n} \slashed{e}_q \otimes \gamma_{\mu_n} \dots \gamma_{\mu_1} \slashed{e}_q \} \right),
\end{equation}
which is simply isomorphic to $\mathbb{C}^{4(n+1)}$. We have defined
\begin{equation}
    \slashed{e}_q \equiv \frac{\slashed{q}}{\sqrt{q^2}}
\end{equation}
for ease of notation.

For completeness, we also define the vector space
\begin{equation}
    \tilde{W}_{n,k} \equiv \left(\mathbb{C}^d\right)^k \oplus \tilde{V}_n,
\end{equation}
whose elements are of the form
\begin{equation}
\begin{split}
    x = \Big( c^{(0)} \slashed{p}_1\dots\slashed{p}_m \otimes \slashed{p}_{m+1}\dots\slashed{p}_k,\, 
    c^{(1)} \slashed{p}_1\dots\slashed{p}_m\, \gamma^\mu\otimes\slashed{p}_{m+1}\dots\slashed{p}_k\, \gamma_\mu,\\
    \sum_i c^{(2)}_i \slashed{p}_1\dots\slashed{p}_m\, \Gamma^{(2)}_i \otimes \slashed{p}_{m+1}\dots\slashed{p}_k\, \Gamma^{(2)}_i,
    \dots\Big),
\end{split}
    \label{eq:n_tuple_slashed}
\end{equation}
in analogy with Eq.~\eqref{eq:n_tuple_gammas}.

\subsection{Construction of projectors}

\subsubsection{General projectors}

The standard way of defining a set of projectors $\{P_i\}$ is to make use of an inner product $\langle \cdot\, , \cdot \rangle$ defined on the corresponding vector space. The projection of $x$ onto a subspace of the vector space is
\begin{align}
P(x) = \sum_i e_i P_i(x) = \sum_{i,j}  e_i \lambda_{ij} \langle e_j, x \rangle,
\end{align}
where the basis elements of the subspace are denoted by $e_i$. The coefficient matrix $\lambda_{ij}$ of the projectors can be determined from demanding that the projector maps basis elements onto themselves,
\begin{equation}
P_i(e_k) =  \sum_j \lambda_{ij} \langle e_j, e_k \rangle = \sum_j \lambda_{ij} \mathsf{G}_{jk} \overset{!}{=} \delta_{ik},
\end{equation}
where we have defined the Gram matrix $\mathsf{G}_{ij} \equiv \langle e_i, e_j \rangle$. The coefficient matrix is simply given by its inverse,
\begin{equation}
\lambda_{ij} = \mathsf{G}_{ij}^{-1}.
\end{equation}
The Gram matrix is invertible if and only if the set of vectors is linearly independent, which is trivially satisfied if we have chosen a proper basis. 

We are thus left with finding an appropriate inner product $\langle \cdot\, , \cdot \rangle: V \times V \rightarrow \mathbb{C}$ which has the following properties: 
\begin{enumerate}[label=(\roman*)]
    \item Conjugation symmetry: $\overline{\langle x, y\rangle} = \langle y, x \rangle$
    \item Linearity: $\langle a x + b y, z\rangle = a \langle x,z\rangle + b \langle y,z\rangle \, \forall\, a,b\in \mathbb{C}$
    \item Positive-definiteness: $\langle x,x\rangle > 0$ 
\end{enumerate}
Note that a projector constructed from an inner product is guaranteed to yield the expected results, but we can (and will) sacrifice property (iii) as long as we explicitly check that the Gram matrix is invertible.

\subsubsection{The standard approach}

To illustrate the above, consider the standard choice for an inner product (see, e.g., Refs.~\cite{Peraro:2020sfm,Tancredi:2022iyo}) on the vector space $W_{n,k}$. For $x,y \in W_{n,k}$, each consisting of a single term, this is
\begin{equation}
\begin{split}
    \langle x, y \rangle \equiv\, &\text{Tr}\bigg[\left(\slashed{p}_1\dots\slashed{p}_{m_x}\,\gamma^{\mu_1} \dots \gamma^{\mu_{n_x}}\right)^\dagger \slashed{p}_1\dots\slashed{p}_{m_y}\,\gamma^{\nu_1} \dots \gamma^{\nu_{n_y}}\bigg] \times \\
    &\text{Tr}\left[\left(\slashed{p}_{m_x+1}\dots\slashed{p}_{k_x}\,\gamma_{\mu_{\sigma(1)}} \dots \gamma_{\mu_{\sigma(n_x)}}\right)^\dagger \slashed{p}_{m_y+1}\dots\slashed{p}_{k_y}\, \gamma^{\nu_{\sigma'(1)}} \dots \gamma^{\nu_{\sigma'(n_y)}}\right],
    \label{eq:standard_inner_product}
\end{split}
\end{equation}

where the indices $\mu_i$ and $\nu_i$ stem from $x$ and $y$ respectively. For property (i) we can choose complex conjugation or pick the interchange of the two spin lines as our conjugation property. In the case of complex conjugation, this property is fulfilled by Eq.~\eqref{eq:standard_inner_product}, but there is no reason why we would need this particular symmetry. The more relevant symmetry is the conjugation where the spin lines are interchanged, which is obeyed by basically anything we would write down intuitively. 

Property (ii) is tricky when the elements have more than one term, e.g.~for $x=\gamma^\mu\otimes\gamma_\mu + \gamma^\mu\gamma^\nu\otimes\gamma_\mu\gamma_\nu$ we would naively get the inner product
\begin{equation}
    \langle x, y \rangle =\,\text{Tr}\Big [(\gamma^\mu + \gamma^\mu \gamma^\nu)^\dagger \dots \Big ] \times\,\text{Tr}\Big [(\gamma_\mu + \gamma_\mu \gamma_\nu)^\dagger \dots\Big ],
\end{equation}
which is obviously inconsistent as the Lorentz indices are not contracted on all terms. The correct way to handle this is to introduce the scalar product to act on each component of the vector element defined in Eq.~\eqref{eq:n_tuple_slashed}:
\begin{equation}
    \langle \cdot\, , \cdot \rangle:  \Big(\left(x_0, x_1,\, \dots\,, x_n\right),\,  \left(y_0, y_1,\, \dots\,, y_n\right)\Big) \mapsto \phi(x_0, y_0) +  \phi(x_1, y_1) + \dots +  \phi(x_n, y_n),
\end{equation}
where the vector index refers to the number of pure $\gamma$ matrices. The map that one might sloppily call the inner product in the definition from Eq.~\eqref{eq:standard_inner_product} is a special case of the scalar map $\phi$,
\begin{equation}
\begin{split}
\phi_t: (\mathbb{C}^d)^m \oplus \tilde{V}^{(k)} \times (\mathbb{C}^d)^m \oplus \tilde{V}^{(k)} &\to \mathbb{C}\\
(x_1 \otimes x_2, \, y_1 \otimes y_2) &\mapsto \text{Tr} \left[ x_1^\dagger y_1\right] \times \text{Tr}\left[ x_2^\dagger y_2 \right]
\end{split}
\label{eq:standard_phi}
\end{equation}
Note, however, that any $\phi$ alone does not obey linearity and hence does not constitute an inner product. The conjugation property (i) of the inner product is inherited from $\phi$ while linearity is explicitly enforced for different lengths of spin lines. For every spin line length, $\phi$ still needs to be linear for (ii) to be fulfilled, but this is something that we can now make sense of and which works for Eq.~\eqref{eq:standard_phi}. Therefore it is important to differentiate between the scalar map $\phi$ which takes elements from $\tilde{V}^{(k)} \times \tilde{V}^{(k)}$ (or $(\mathbb{C}^d)^m \oplus \tilde{V}^{(k)} \times (\mathbb{C}^d)^m \oplus \tilde{V}^{(k)}$ when considering slashed momenta) and the inner product $\langle \cdot\, , \cdot \rangle$ which takes elements from $\tilde{V}_n \times \tilde{V}_n$ (or $\tilde{W}_{m,n} \times \tilde{W}_{m,n}$ when considering slashed momenta). 

Finally, with the traditional map $\phi_t$, the inner product also obeys property (iii), but as we will see in Section \ref{sec:new_phi}, this is less important.

\subsubsection{An alternative path}
\label{sec:new_phi}

One less appealing aspect of the standard map $\phi_t$, which is used to build up the inner product, is that all Lorentz indices are contracted across the two spin lines. In automated computations this means that the intermediate expressions may become quite large once we start taking one of the traces. Moreover, there is no obvious way of parallelising such a calculation. A trace calculation arising in the projection of 9 or more $\gamma$ matrices on either spin line runs multiple days on a single core even with a highly optimised programming language like \texttt{FORM} \cite{Kuipers:2012rf}. A possible solution is to choose a different map $\phi$. For this purpose we define
\begin{equation}
\begin{split}
    \phi_a: (\mathbb{C}^d)^m \oplus \tilde{V}^{(k)} \times (\mathbb{C}^d)^m \oplus \tilde{V}^{(k)} &\to \mathbb{C} \\
    (x_1 \otimes x_2, \, y_1 \otimes y_2) &\mapsto \text{Tr} \left[ x_1 y_1 x_2 y_2 \right]
    \label{eq:map_single_trace}
\end{split} 
\end{equation}
This effectively glues together the two spin lines, which has the nice feature that all Lorentz indices are contracted on the same spin line. This allows for two $\gamma$ matrices to be commuted together and eliminated from the spin line, thereby generating many terms which have, however, two $\gamma$ matrices less per index contraction that has been carried out. As a consequence, evaluating the inner product and carrying out the trace can easily be parallelised.

It is instructive to check the properties for the inner product (i), (ii) and (iii) in this case too. We can see that linearity is obeyed and, with the interchange of the two spin lines as our conjugation of choice, property (ii) also holds. However, this map does not obey positive-definiteness (e.g.~for $x=\gamma^\mu \otimes \gamma_\mu$). This can become a problem when inverting the Gram matrix. 

Focusing on the subspace of pure $\gamma$ matrices only for ease of notation, positive-definiteness over $V_n$ (but not necessarily $\tilde{V}_n$) ensures that the Gram matrix is invertible as long as the set of vectors considered is a basis, i.e.~linearly independent. 
\begin{equation}
\langle x, x \rangle > 0 \,\, \forall \,\, x \in V_n : x \neq 0 \iff \mathbf{v}^\dagger \mathsf{G} \mathbf{v} > 0 \,\, \forall \,\, \mathbf{v} \in \mathbb{C}^n \cong V_n : \mathbf{v} \neq 0, 
\end{equation}
where $\mathbf{v}$ represents an element in $V_n$, containing just the coefficients in front of the Dirac structures in the $n$-tupel. If the inner product is not positive definite, we can have eigenvectors with zero eigenvalue, making the matrix non-invertible:
\begin{align}
\langle x, x \rangle > 0 & \implies \forall \mathbf{v} \neq 0: \mathsf{G} \mathbf{v} \neq 0, \\
\exists \mathbf{v}: \mathsf{G} \mathbf{v} = 0 &\iff \mathsf{G} \,\,\, \text{non-invertible}.
\end{align}
It turns out that this is not a problem for the vector space of just $\gamma$ matrices, but if we include insertions of the external momentum $\slashed{q}$ as well, the Gram matrix becomes non-invertible in some cases.

We will describe the construction of our basis and how we can avoid non-invertibility of the Gram matrix in the following. When projecting the Dirac structures occurring in the amplitude, we count the number $n$ of $\gamma$ matrices on the shorter spin line and include projectors up to that number of pure $\gamma$ matrices and at most one more slashed momentum, which corresponds to the most complicated structure that can occur. If the longer of the two spin lines has more than $n+1$ $\gamma$ matrices, there must be index contractions on that spin line, reducing the number of $\gamma$ matrices down to at most $n+1$. If $n$ is even, the Gram matrix constructed in this way is non-invertible. This can be alleviated by either extending the basis to include one more $\gamma$ matrix on each spin line or by excluding the basis element with $n + n$ pure $\gamma$ matrices and an additional slashed momentum on the spin line which has only $n$ $\gamma$ matrices in the amplitude. We choose the latter option which makes the Gram matrix invertible and is the computationally most efficient approach.

The case where both spin lines have the same length of eleven $\gamma$ matrices can be considered separately in our case because these diagrams have ten pure $\gamma$ matrices and one slashed momentum per spin line. Since extending the basis to twelve $\gamma$ matrices is computationally expensive, we instead choose a different basis as the one described above. For this specific symmetric case, the basis includes all basis elements with up to eleven $\gamma$ matrices on each spin line (including slashed momenta). However, to make the Gram matrix invertible we need to exclude one asymmetric basis element which has ten $\gamma$ matrices on both spin lines and a slashed momentum on one of the two. For more details, see Appendix \ref{app:basis}.

\subsection{An efficient treatment of pure \texorpdfstring{\boldmath$\gamma$}{gamma} matrices}
\label{sec:pure_gammas_permutation}

For spin lines which consist only of pure $\gamma$ matrices, the problem is reduced to finding the maps
\begin{equation}
f_k: S_k \rightarrow \mathbb{R}^{k+1},
\end{equation}
where $S_k$ is the group of permutations of $k$ elements. The function $f_k$ take a permutation $\sigma$ of Lorentz indices, e.g.~of the second spin line if the first has been labelled canonically, and returns the coefficients $\mathbf{a}$ of the ordered spinor structures. For later convenience we choose $a_1$ to correspond to the coefficient of the longest Dirac chain, whereas the coefficient $a_{k+1}$ for each $f_k$ corresponds to the coefficient of the identity. Hence, the function $f_k$ resolves the mapping onto the vector space defined in Eq.~\eqref{eq:gamma_basis}. The first few mappings can be written down by commuting the $\gamma$ matrices on the second spin line into canonical order:
\begin{equation}
f_1: S_1 \rightarrow \mathbb{R}^2, \quad (1) \to (1,0)
\end{equation}
\begin{equation}
f_2: S_2 \rightarrow \mathbb{R}^3, \quad \begin{cases}
(12)\mapsto (1,0,0)\\
(21)\mapsto (-1,0,2d)
\end{cases}\label{eq:perm_2}
\end{equation}
\begin{equation}
f_3: S_3 \rightarrow \mathbb{R}^4, \quad \begin{cases}
(123)\mapsto (1,0,0)\\
(132)\mapsto (-1,0,2d,0)\\
(231)\mapsto (1,0,4-4d,0)\\
\quad\vdots
\end{cases}
\end{equation}
The permutation notation $(\dots)$ can be read as the numbering of the Lorentz indices $\mu_i$ on the second spin line from back to front. The unpermuted elements are the basis vectors which span the vector space in Eq.~\eqref{eq:gamma_basis}. For example, the non-trivial permutation in Eq.~\eqref{eq:perm_2} translates to
\begin{equation}
\gamma^{\mu_1} \gamma^{\mu_2} \otimes \gamma_{\mu_1} \gamma_{\mu_2} = - \gamma^{\mu_1} \gamma^{\mu_2} \otimes \gamma_{\mu_2} \gamma_{\mu_1} + 2d( \mathds{1} \otimes \mathds{1}).
\end{equation}
From the above we can already see a pattern emerge. For example, the first coefficient, which corresponds to the contribution from the ordered spin line of the same length $k$, is always the sign of the permutation, i.e.~$a_0=\text{sgn}(\sigma)$. Moreover, the odd coefficients $a_{2n+1}$ always vanish; this is the case since any index contraction can only reduce the number of $\gamma$ matrices by two on each spin line. Closed formulae for the above expressions are tricky to obtain, but generating a lookup table for all permutations of up to a number $k=10$ of $\gamma$ matrices is feasible and took $\mathcal{O}(10^6)$ single core CPU minutes.

This treatment of the pure $\gamma$ matrices implements the mapping of pure gamma matrices onto a minimal basis described as in Section \ref{sec:pure_vector_spaces} and considerably reduces the effort needed to project the encountered spinor structures in our amplitude. This is because we only need to project a limited number of momentum insertions paired with a certain number of pure $\gamma$ matrices.

\subsection{Implemented algorithm}
\label{sec:algorithm}

\begin{figure}
    \centering
    \includegraphics[scale=0.6]{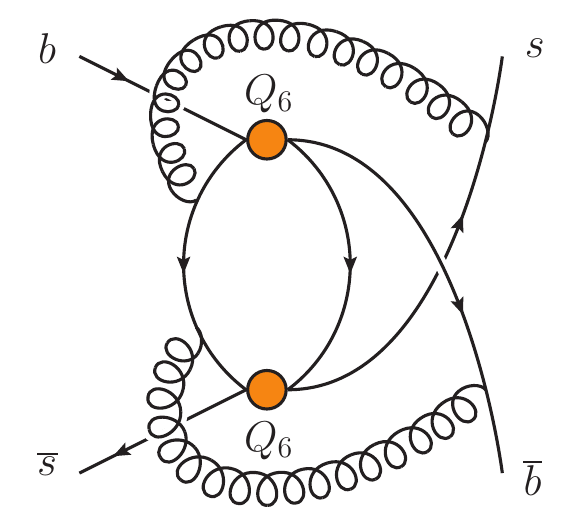}
    \caption{The spinor structure of the diagram shown here will be used as an example to illustrate how our algorithm works on the most complicated diagrams. Note that the Lorentz indices corresponding to the effective operators $Q_6$ are labelled as $\mu_i$ and $\nu_i$ while the gluons carry Lorentz indices $\alpha_i$.}
    \label{fig:nonplanar_3l}
\end{figure}

With the general background outlined in the preceding sections, we are ready to outline the algorithm chosen to deal with the $\gamma$ matrices in our amplitude. The goal is to map each diagram onto a sum of operator matrix elements in the $\lvert\Delta B \rvert = 2$ theory. This is done in two steps; first, the $\gamma$ matrices are ordered canonically and then the projectors are applied. The full procedure is given below and for illustration purposes we will refer to the structures appearing at each step of the calculation of the diagram shown in Fig.~\ref{fig:nonplanar_3l}.
\begin{enumerate}[label=(\roman*)]
\item Project onto left-handed and right-handed spinor structures. It is sufficient to just multiply with $P_{R/L} = (1\pm\gamma_5)/2$ and then drop all remaining terms containing $\gamma_5$. None of these can contribute to the amplitude because the left/right projectors can always be commuted through all the way to the end of a spin line and no $\gamma_5$ should appear as an overall factor. For the sample diagram, we have
\begin{equation}
    \gamma^{\mu_1} \gamma^{\mu_2} \gamma^{\mu_3} \slashed{k}_1 \gamma^{\alpha_1} \slashed{k}_2 \gamma^{\alpha_2} \slashed{k}_ 3\gamma^{\nu_1} \gamma^{\nu_2} \gamma^{\nu_3} \otimes \gamma_{\alpha_2} \slashed{k}_4 \gamma_{\mu_1} \gamma_{\mu_2} \gamma_{\mu_3} \slashed{k}_5 \gamma_{\nu_1} \gamma_{\nu_2} \gamma_{\nu_3} \slashed{k}_6 \gamma_{\alpha_1} + \text{3 terms}\,,
\end{equation}
where the $k_i$ are line momenta of the Feynman diagram in Fig.~\ref{fig:nonplanar_3l}.
\item Express all line momenta in terms of loop momenta and the external momentum, which reduces the number of slashed $\gamma$ matrices which may appear, e.g.~for the sample diagram:
\begin{equation}
    \gamma^{\mu_1} \gamma^{\mu_2} \gamma^{\mu_3} \slashed{p}_3 \gamma^{\alpha_1} \slashed{p}_3 \gamma^{\alpha_2} \slashed{p}_2 \gamma^{\nu_1} \gamma^{\nu_2} \gamma^{\nu_3} \otimes \gamma_{\alpha_2} \slashed{p}_3 \gamma_{\mu_1} \gamma_{\mu_2} \gamma_{\mu_3} \slashed{p}_1 \gamma_{\nu_1} \gamma_{\nu_2} \gamma_{\nu_3} \slashed{p}_2 \gamma_{\alpha_1} + \text{95 terms}\,.
\end{equation}
\item Reorder the momenta on both spin lines:
\begin{enumerate}
    \item Convert duplicate momenta on the same spin line to scalar products. This generates at most a factor of ten more terms, and in the example considered here, we obtain 419 terms after this step. %<1s computation time up to here
    \item Commute all slashed momenta to the left of the spin lines. This generates at most a factor of 1000 more terms.
    \item All occurrences of the external momentum are put to the very left, followed by the loop momenta. Putting the external momentum to the left of the spin lines is advantageous since some of the basis elements contain a slashed external momentum which is always on the right. Applying such a projector immediately gives a scalar product, thereby reducing the length of the trace. In the sample diagram we have 318808 terms after this step. %about 330s single core computation time up to here
    \item Choose one particular permutation of the loop momenta.
\end{enumerate}
After reordering all momenta, the sample diagram has the following spinor structure:
\begin{equation}
    \slashed{p}_3 \slashed{p}_2 \slashed{p}_1 \gamma^{\mu_2} \gamma^{\mu_3} \gamma^{\alpha_2} \gamma^{\nu_1} \gamma^{\nu_2} \gamma^{\nu_3} \otimes \slashed{p}_3 \gamma_{\alpha_2} \gamma_{\mu_2}\gamma_{\mu_3} \gamma_{\nu_1} \gamma_{\nu_2} \gamma_{\nu_3} + \text{181510 terms}
\end{equation}
\item Contract duplicate Lorentz indices on the same spin line after commuting the respective $\gamma$ matrices together.
\item Bring all pure $\gamma$ matrices into a canonical order. We use a lookup table which resolves the permutation of pure $\gamma$ matrices, see Section \ref{sec:pure_gammas_permutation}. This table contains all the mappings of permutations of up to ten $\gamma$ matrices on both spin lines onto linear combinations of canonically ordered $\gamma$ matrices. To generate less terms when using the map $\phi_a$ from Eq.~\eqref{eq:map_single_trace} to carry out the projection, it will be useful to choose a canonical ordering where the $\gamma$ matrices on the two spin lines are inverted with respect to each other. This way the largest number of necessary commutations is given by the length of the projector element which is sandwiched between the spin lines. For our example case, this leads to
\begin{equation}
    \slashed{p}_3 \slashed{p}_2 \slashed{p}_1 \gamma^{\rho_1} \gamma^{\rho_2} \gamma^{\rho_3} \gamma^{\rho_4} \gamma^{\rho_5} \gamma^{\rho_6} \otimes \slashed{p}_3 \gamma_{\rho_6} \gamma_{\rho_5}\gamma_{\rho_4} \gamma_{\rho_3} \gamma_{\rho_2} \gamma_{\rho_1}  + \text{40134 terms}.
    %700s up to here, but didn't optimise by bracketing or putting scalar products in wrappers
\end{equation}
\item Map the terms with ordered pure $\gamma$ matrices and split-off slashed momenta directly onto basis elements which are given in Appendix \ref{app:basis}. This is done by using another lookup table which has been calculated using projectors on the ordered gamma structures. Note that the process is symmetric under interchange of the spin lines; hence, it is only necessary to calculate one of the two possibilities whenever the loop momenta distribution is asymmetric with respect to interchange of the spin lines. When calculating the lookup table, we can also choose to use different projectors, i.e.~different scalar products, depending on the spinor structure which needs to be resolved. For most structures, we choose to project using the bilinear map $\phi_a$ defined in Eq.~\eqref{eq:map_single_trace}. For the projection of the spinor structure with eleven $\gamma$ matrices and slashed momenta on both spin lines, the aforementioned linear map is no longer efficient since the slashed momenta are essentially open Lorentz indices inside the trace. Thus we use the linear map $\phi_t$ which takes the traces over both spin lines separately as defined in Eq.~\ref{eq:standard_inner_product}. To optimise this calculation, the terms are split into separate files in bunches of $10^5$ terms after taking the first trace. The parallelisation of the second trace makes this calculation tractable, although the size of the intermediate files generated is also a challenge. In the end, our sample diagram results in 
\begin{equation}
    \frac{(p_1\cdot p_3)^2 p_2^2}{d^3 - 6d ^2 + 11 d - 6} \times B_{45}  + \text{104335 terms},
\end{equation}%720s up to here
where $d=4-2\epsilon$ is the spacetime dimension, and $B_{45}$ is defined in Appendix \ref{app:basis}.
\item Map basis elements onto operator matrix elements, e.g.~$\langle B_s|Q|\bar{B}_s \rangle$, $\langle B_s|\widetilde{Q}_S|\bar{B}_s \rangle$, $\langle B_s| E^{(1)}_1|\bar{B}_s \rangle$, etc. This can be done using a lookup table.
\end{enumerate}
The algorithm described above is well suited for an efficient calculation since it circumvents a few bottlenecks such as:
\begin{itemize}
\item No redundant calculation of long traces. Each trace combination is only taken once.
\item Applying the projectors by multiplying the amplitude with them inflates the number of terms by  one or two orders of magnitude if the different chirality combinations are taken into account. This is avoided by using a lookup table and mapping directly onto the result. 
\item The number of spinor structures which need to be projected is massively reduced by the hybrid approach of sorting the $\gamma$ matrices first.
\end{itemize}

%- }}}
%- {{{ Tensor integrals:

\section{\label{sec:tensor}Tensor integrals}

Tensor reduction represents an alternative approach to the calculation of amplitudes on both sides of the matching equation. In comparison to the projector technique it does not require
any assumptions on the Dirac and colour structures that may appear in the resulting expressions.  Furthermore, when using this method, evanescent operators containing large numbers of $\gamma$ matrices do not lead to any performance bottlenecks.

However, tensor reduction formulas tend to become very large with the increasing tensor rank (number of open Lorentz indices), number of loop momenta (more ways to permute the indices) and the number of external momenta (more complicated tensor structures). In our calculation we need
to consider three-loop tensors up to rank 11 with one external momentum.

The traditional way of deriving tensor reduction formulas consists of writing down the most general ansatz containing all tensor structures allowed by the symmetries. For example, 
in the case of a rank-three two-loop integral with one external momentum $q$, one can naively write

\begin{align}
\int_{p1,p2} p_1^\mu p_2^\nu p_2^\rho f(p_1,p_2,q) = 
\int_{p1,p2} \left ( g^{\mu \nu} q^\rho c_1 + g^{\mu \rho} q^\nu c_2  + g^{\nu \rho} q^\mu c_3 +
q^\mu q^\nu q^\rho c_4 \right )  f(p_1,p_2,q),
\label{eq:tdec}
\end{align}
where $f(p_1,p_2,q)$ is some function containing propagator denominators and possibly scalar 
products appearing in the numerator of this loop integral. Its precise form is irrelevant for the actual tensor reduction. The coefficients $c_i$ are made of scalar products containing both loop and external momenta multiplied by rational functions that depend on the dimension $d$.

The next step would be to contract Eq.~\eqref{eq:tdec} with each of the tensor structures multiplying $c_i$, which leads to a symbolic system of linear equations. Solving this system we can determine the $c_i$, thus deriving the final form of the tensor reduction formula for this
integral. 

In practical applications one quickly arrives at very large systems of equations that are barely solvable in a reasonable amount of time without special codes. The complexity of such equations can be greatly reduced by observing that not all $c_i$ are independent of each other. A very simple and efficient algorithm for finding such symmetries is
described in Ref.~\cite{Pak:2011xt}. It is enough to contract the \emph{left-hand side} of Eq.~\eqref{eq:tdec} with each of the tensor structures multiplying $c_i$ and compare the resulting expressions with each other. Identical expressions mean that the corresponding $c_i$ are the same so that there are symmetry relations that reduce the size of the system.

The whole procedure of deriving reduction formulas for arbitrary tensor integrals can be automatized using software tools. In particular, one can use the routine \texttt{Tdec} available in \textsc{FeynCalc} 10 \cite{Mertig:1990an,Shtabovenko:2016sxi,Shtabovenko:2020gxv,Shtabovenko:2023idz}, which already implements the symmetry finding algorithm from Ref.~\cite{Pak:2011xt}. In order to speed up the process of solving the linear system, it is also recommended to install the development version of the \textsc{FeynHelpers}\footnote{\url{https://github.com/FeynCalc/feynhelpers}} add-on \cite{Shtabovenko:2016whf}.
This code provides an interface to the computer algebra system \textsc{Fermat} \cite{Lewis:Fermat} that features a very efficient linear equation solver. To make use of this, one needs to set the option \texttt{Solve} of \texttt{Tdec} to \texttt{FerSolve}.

The full \textsc{Mathematica} code for deriving the reduction formula for the integral from Eq.~\eqref{eq:tdec} reads as follows

\begin{verbatim}
$LoadAddOns = {"FeynHelpers"};
<< FeynCalc`
Tdec[{{p1, mu}, {p2, nu}, {p2, rho}}, {q}, Solve -> FerSolve, 
 List -> False, FCE->True]
\end{verbatim}

This produces the following output
\begin{verbatim}
(FVD[q, rho]*MTD[mu, nu]*SPD[p2, q]*(-(SPD[p1, q]*SPD[p2, q]) + 
SPD[p1, p2]*SPD[q, q]))/((-1 + D)*SPD[q, q]^2) +  
(FVD[q, nu]*MTD[mu, rho]*SPD[p2, q]*(-(SPD[p1, q]*SPD[p2, q]) + 
SPD[p1, p2]*SPD[q, q]))/((-1 + D)*SPD[q, q]^2) + 
(FVD[q, mu]*MTD[nu, rho]*SPD[p1, q]*(-SPD[p2, q]^2 + 
SPD[p2, p2]*SPD[q, q]))/ ((-1 + D)*SPD[q, q]^2) + 
(FVD[q, mu]*FVD[q, nu]*FVD[q, rho]*(2*SPD[p1, q]*SPD[p2, q]^2 + 
D*SPD[p1, q]*SPD[p2, q]^2 - SPD[p1, q]*SPD[p2, p2]*SPD[q, q] 
- 2*SPD[p1, p2]*SPD[p2, q]*SPD[q, q]))/((-1 + D)*SPD[q, q]^3)
\end{verbatim}
where \texttt{SPD} denotes $d$-dimensional scalar products of four-momenta, \texttt{MTD} stands for the metric tensor and \texttt{FVD} describes four-vectors. Using string replacements such expressions can be easily converted into \textsc{FORM} \texttt{id}-statements and thus used outside of \textsc{Mathematica}. 

The insertion of such \texttt{id}-statements into amplitudes can significantly increase the number of intermediate terms, so that putting \texttt{.sort} commands in right places of the \textsc{FORM} code becomes mandatory. However, the derivation of such formulas using \textsc{FeynCalc} and \textsc{FERMAT} was never a performance bottleneck for us. For example, to derive tensor reduction for

\begin{align}
\int_{p1,p2,p3} p_1^{\mu_1} p_1^{\mu_2} p_1^{\mu_3} p_1^{\mu_4} p_1^{\mu_5}
p_2^{\mu_6} p_2^{\mu_7} p_2^{\mu_8} p_2^{\mu_9} p_2^{\mu_{10}} p_2^{\mu_{11}} f(p_1,p_2, p_3,q)
\end{align}

we needed only 3 minutes on a modern laptop, while the resulting \texttt{id}-statement saved as a text file had the size of 2.8 MB.

%- }}}
%- {{{ Master integrals:

\section{\label{sec:master}Master integrals}

In this Section we describe our approach for the computation of the master
integrals.

To obtain results published in Refs.~\cite{Gerlach:2022wgb,Gerlach:2022hoj} it was sufficient to perform a naive Taylor expansion of the $|\Delta B| = 1$ amplitudes in $z$ up to $\mathcal{O}(z)$. Therefore, we were only concerned with the calculation of master integrals in the $z\to0$ limit. Analytic results for one- and two-loop integrals can be readily found in the literature \cite{Smirnov:2012gma,Fleischer:1999tu} but most of the three-loop integrals had to be calculated from scratch.
 We start with the description of these analytic calculations in Section~\ref{sub::ana}.

After the publication of Refs.~\cite{Gerlach:2022wgb,Gerlach:2022hoj} we were able to extend our two-loop results to higher orders in $z$ by keeping the full $z$-dependence during the evaluation of the amplitudes and expanding the resulting master integrals in $z$ asymptotically. The technicalities behind this method are described in Section~\ref{sub::ana-asy}.

Afterwards, in Section~\ref{sub::semi} we describe a semi-numerical method we use to obtain results
to very high orders in $z$ for all relevant integrals at two and three loops. In particular, it was important to cover the physical range for all possible numerical values of the quark masses. Depending on the renormalization scheme, the charm and bottom quark masses
typically take values between $1-1.8$~GeV and $4.2-5.0$~GeV, respectively. Allowing for
different renormalization schemes of the quark masses, this leads to values for
$\sqrt{z}$ between $0.20$ and $0.43$.

\subsection{\label{sub::ana}Analytic calculation for \texorpdfstring{\boldmath$z\to0$}{z to 0}}

    In the $z \to 0$ limit, the final three-loop master integrals  can be calculated using the direct integration of their Feynman parametric representations. In the following we will describe how these calculations can be streamlined using  the \textsc{Mathematica} package \textsc{FeynCalc} and the \textsc{Maple} package \textsc{HyperInt} \cite{Panzer:2014caa}.

The first step is to derive the Feynman parametric representation for each of the integrals,
where we set $m_b$ to $1$ as it can be recovered later using dimensional analysis. Here we make use of the routine \texttt{FCFeynmanParametrize} that is 
available in \textsc{FeynCalc} 10 and can automatically generate such
parametrizations for a wide range of multiloop integrals. 

Given an integrand \texttt{int} in the propagator representation, e.g.
\begin{verbatim}
int = FAD[p1]*FAD[p2]*FAD[p1 - p2 - p3]*FAD[p1 + q]*FAD[p3 + q]*
FAD[p2 + p3 + q]*FAD[{p1 - p2, mb}]*FAD[{p3, mb}];
\end{verbatim}
it is sufficient to evaluate\footnote{The options \texttt{Names} and \texttt{Indexed} specify that the Feynman parameters should be of the form \texttt{x[i]}, while \texttt{Simplify} and \texttt{Assumptions} allow for some simplifications of the resulting integrand. With \texttt{FCReplaceD} we can replace the number of dimensions $d$ in favor of $4-2\epsilon$, while \texttt{FinalSubstitutions} is used to replace the occurring kinematic invariants with explicit values. More information on this and other \textsc{FeynCalc}  functions can be found in the official PDF manual available at \url{https://github.com/FeynCalc/feyncalc-manual/releases/tag/dev-manual}.}
\begin{verbatim}
FCFeynmanParametrize[int, {p1, p2, p3}, Names -> x, 
  Indexed -> True, FCReplaceD -> {D -> 4 - 2 ep}, Simplify -> True,
  Assumptions -> {ep > 0}, FinalSubstitutions -> 
  {SPD[q] -> 1, mb^2 -> 1}]
\end{verbatim}
which returns a list of three entries. These are the integrand itself, its
prefactor independent of the integration variables (the piece mostly made of $\Gamma$-functions) and the list of  the Feynman parameters $x_i$. As we will also need the Symanzik 
polynomials $\mathcal{U}$ and $\mathcal{F}$ separately, we need to
run\footnote{\textsc{HyperInt} examples and documentation commonly use $\psi$ and $\phi$ to denote the Symanzik polynomials $\mathcal{U}$ and $\mathcal{F}$ respectively.}
\begin{verbatim}
{psi, phi} = 
 FCFeynmanPrepare[int, {p1, p2, p3}, Names -> x, Indexed -> True, 
   FinalSubstitutions -> {SPD[q] -> 1, mb^2 -> 1}][[1 ;; 2]]
\end{verbatim}
Even though we are working with single-scale integrals, they involve
up to 8 integrations in Feynman parameters and, furthermore, can be divergent. Therefore, trying to solve such integrals analytically using \textsc{Mathematica} alone is not feasible. To deal with these issues we make use of \textsc{HyperInt}, a \textsc{Maple} package that was developed with such calculations in mind.

When evaluating an integral with \textsc{HyperInt}, one usually makes use
of the Cheng-Wu theorem \cite{Cheng:1987ga} which implies that the given integral
must be projective. The projectivity property of a Feynman parametric integral 
means that the integrand must remain invariant under the rescaling
\begin{equation}
	x_i \to \lambda x_i, \quad d x_i \to \lambda d x_i,
\end{equation} 
so that $\lambda$ must explicitly cancel out. In \textsc{FeynCalc} the projectivity can be checked using the routine \texttt{FCFeynmanProjectiveQ}.

The practical application of Cheng-Wu comes from the statement 
that if we have a projective integral, then the Dirac delta in its definition can be replaced with
\begin{equation}
	\delta \left ( \sum_{i \in \sigma} x_i -1 \right )
\end{equation} 
without changing the result, where $\sigma$ is an arbitrary subset of the edges. Although the choice of $\sigma$ can be highly nontrivial, in practice one often opts for setting a single Feynman parameter, say $x_k$, to unity. The remaining Feynman parameters $x_i$ with $i \neq k$ are then integrated from $0$ to $\infty$.

When using \textsc{HyperInt}, a Feynman parametric integral is calculated order by order in~$\epsilon$. Expanding a divergent integral in the regulator will introduce divergences in the integrations over Feynman parameters, which one normally would like to avoid. To this end, \textsc{HyperInt} implements the method of analytic regularization \cite{Panzer:2014caa,Panzer:2014gra,Panzer:2015ida} where the unexpanded integrand is modified in such a way, that all divergences will show up as explicit  $\epsilon$-poles, while the integrations over $x_i$ become manifestly finite. The corresponding \textsc{HyperInt} 
routines are called \texttt{findDivergences} and \texttt{dimregPartial}, while their \textsc{FeynCalc} counterparts (using the same algorithm as in \textsc{HyperInt}) go under the names of \texttt{FCFeynmanFindDivergences} and \texttt{FCFeynmanRegularizeDivergence}.

Unfortunately, this technique also tends to increase the complexity of the integrand, especially if it contains a lot of divergences that need to be regularized. This is why in practice people often choose a different method called \textit{quasi-finite basis} \cite{vonManteuffel:2014qoa}. However, for the integrals at hand, this was not a major issue and with the exception of one 5-edge integral we kept the original basis and naively used analytic regularization.

When trying to integrate the regularized integrals using \textsc{HyperInt}'s \texttt{hyperInt} function, we found that in most cases the process could not be finished due to the appearance of spurious divergences and nonlinear polynomials at intermediate stages. Following the advice of the package developer\footnote{Private communication with Erik Panzer.} we were able to resolve these issues by setting the options \texttt{_hyper_abort_on_divergence} and \texttt{_hyper_ignore_nonlinear_polynomials} to \texttt{false} for all but the very last integration in $x_i$. In the case of the last integration these options must be set to \texttt{true}, while the option \texttt{_hyper_splitting_field} must be e.g.~set to 
\begin{verbatim}
{RootOf(_Z^2 + 3*_Z + 1),RootOf(_Z^2 + _Z + 1),RootOf(_Z^2 + 1)};
\end{verbatim}
to allow for factorizing nonlinear polynomials using complex numbers.

Upon combining the results obtained at each order in $\epsilon$, we still
need to face two issues. First of all, in some cases \textsc{HyperInt} has
to integrate over the branch cut and cannot automatically determine the sign of the imaginary
part. Such terms contain a $\delta_{x_i}$ which can, in principle, be
$+1$ or $-1$. Consequently, a $\delta_{x_i}^2$ can be readily set to unity.
An easy way to fix these signs consists of evaluating the same integral
numerically using \textsc{pySecDec} \cite{Borowka:2017idc,Borowka:2018goh,Heinrich:2021dbf,Heinrich:2023til} or \textsc{FIESTA} \cite{Smirnov:2015mct,Smirnov:2021rhf}
at several points and comparing the outcome to the analytic result, where
everything apart from the $\delta_{x_i}$'s is numerical as well. For the cases
at hand we had no difficulties fixing those signs unambiguously.

Second, the analytic results obtained by \textsc{HyperInt} upon using the
routine \texttt{hyperInt} are usually not written in the most compact way.
The standard procedure of simplifying them with the aid of the function
\texttt{fibrationBasis} fails for most of our integrals. This is because
they depend on Goncharov Polylogarithms (GPLs)~\cite{Goncharov:1998kja} containing 6th root of unity,
a class of functions that is, as of now, not included in the simplification
tables shipped with \textsc{HyperInt}. Following the advice of the 
\textsc{HyperInt} developer we used the package \textsc{HyperLogProcedures} 
\cite{Schnetz:HLP}
In most cases the command \texttt{Convert(expr,f23)}\footnote{
In \textsc{HyperLogProcedures} \texttt{f23} denotes a particular alphabet related to
Multiple Deligne values that were studied in Ref.~\cite{Broadhurst:2014jda}.} was sufficient to 
 achieve the required simplifications. Some challenges that we faced when
 using the code were resolved with the kind of help  of the package developer\footnote{Private communication with Oliver Schnetz.}.

Owing to the nature of our problem, where we only need the imaginary parts of the
loop integrals but end up calculating the full result, the separation between
real and imaginary parts required some additional work. To that aim we 
loaded the output of \textsc{HyperLogProcedures} into \textsc{Mathematica}
and performed the final simplifications with the aid of \textsc{PolyLogTools} \cite{Duhr:2019tlz}.

\subsection{\label{sub::ana-asy}Asymptotic expansion around \texorpdfstring{{\boldmath $z = 0$}}{z equals 0}}

As far as the two-loop integrals are concerned, analytic results for the expansion
around $z=0$ can be easily obtained using the method of regions \cite{Beneke:1997zp}.
These results were relevant for extending our theoretical predictions from \cite{Gerlach:2022wgb}
beyond $\mathcal{O}(z)$. Here we would like to explain how the technique of asymptotic
expansions can be applied in a highly automatized fashion using publicly available software
tools including \textsc{FeynCalc} 10.

The danger of overlooking a contributing region by having an inappropriate
routing of the external momentum can be avoided by using the program \textsc{asy} \cite{Pak:2010pt,Jantzen:2012mw}.
This code can reliably identify all regions relevant for the given scaling
hierarchy, but its output is not always straightforward to interpret.
While \textsc{asy} assumes that the expansion will be carried out on the level of Feynman
parametric integrals, in practice one would rather like to expand in masses and 
momenta. The latter approach allows us to use integration-by-parts (IBP) reduction so that a calculation
to any order in $z$ can be always reduced to the same set of master integrals.

Let us remark that such an expansion is not always possible for any kind of loop integral. For example, in Soft-Collinear Effective theory (SCET) \cite{Bauer:2000yr,Bauer:2001yt,Beneke:2002ph,Beneke:2002ni} we need to deal with integrals, 
where different components of four-momenta decomposed along the light-cone vectors $n$ and $\bar{n}$
scale in a different way. In that case it is clearly not possible to expand the integral in the original four-momenta. However, in the case at hand such complications do not occur.

To be more specific, let us discuss the analysis of the following two-loop integral with external momentum $q^2=m_b^2$ using \textsc{asy}
\begin{equation}
\int_{p_1,p_2} \frac{1}{-p_1^2} \frac{1}{m_c^2 - (p_1+q)^2} \frac{1}{-p_2^2} \frac{1}{-(p_1+p_2+q)^2},
\end{equation}
where we want to expand in the limit $m_c \ll m_b$.

Using the code 
\begin{verbatim}
Get["asy21.m"]
SetOptions[QHull, Executable -> "/usr/bin/qhull"];
props = {-p1^2, mc^2 - (p1 + q)^2, -p2^2, -(p1 + p2 + q)^2}
res = AlphaRepExpand[{p1, p2},
  props, {q^2 -> mb^2}, {mb -> x^0, mc -> x^1}, Verbose -> False]
\end{verbatim}
we get the output
\begin{align}
&(0, -2, -2, -2), \nonumber  \\
& (0, \phantom{-}0, \phantom{-}0, \phantom{-}0). \label{eq:asy1}
\end{align}
Notice that switching the first two propagators as in
\begin{verbatim}
props = {mc^2 - (p1 + q)^2, -p1^2, -p2^2, -(p1 + p2 + q)^2}
\end{verbatim}
will produce 
\begin{align}
&(0, 0, 0, 0), \nonumber \\
&(0, 2, 0, 0), \label{eq:asy2}
\end{align}
which is in fact equivalent to Eq.~\eqref{eq:asy1}.

The row vectors returned by \textsc{asy} denote the scalings of propagators
in each contributing region, while their number gives us the number of regions. 
The scalings in each row vector are always relative with respect to each other, which allows us to ``normalize'' such a vector by adding the same integer to each of its entries.
Applying this procedure to the first vector in Eq.~\eqref{eq:asy1} leads to the second vector in Eq.~\eqref{eq:asy2}.

To understand the meaning of those scalings, it is convenient to think of the row vectors as of
lists of powers of a quantity $\Lambda$  that describes the scaling of each inverse propagator. Then, the first output corresponds to 
\begin{align}
&(0, -2, -2, -2) \Leftrightarrow (\Lambda^0, \Lambda^{-2}, \Lambda^{-2}, \Lambda^{-2}), \nonumber \\
& (0, \phantom{-}0, \phantom{-}0, \phantom{-}0) \Leftrightarrow (\Lambda^0, \Lambda^0, \Lambda^0, \Lambda^0), 
\end{align}
meaning that in the first region the contribution of $[-p_1^2]$ is much larger than that of the  other three inverse propagators. The second output given by
\begin{align}
&(0, 0, 0, 0)   \Leftrightarrow (\Lambda^0, \Lambda^0, \Lambda^0, \Lambda^0), \nonumber \\
&(0, 2, 0, 0) \Leftrightarrow (\Lambda^0, \Lambda^{2}, \Lambda^{0}, \Lambda^{0}),
\end{align}
which essentially tells us the same thing. This explains why we are allowed to normalize these row vectors. Multiplying each entry by some power of $\Lambda$ obviously will not change the relative scalings between the propagators.

The output
\begin{align}
&(2, 0, 0, 0) \\
& (0, 0, 0, 0),
\end{align}
tells us that the asymptotic expansion of our integral in the limit $m_c \ll m_b$ receives contributions from two regions.

In the first region the inverse propagators scale as 
\begin{equation}
[-p_1^2] \sim m_b^2, \quad [m_c^2 - (p_1 + q)^2] \sim [-p_2^2] \sim [-(p_1 + p_2 + q)^2]  \sim m_c^2.
\end{equation}
Here it is convenient to perform the shift $p_1 \to p_1 - q$, as momentum shifts do not alter the scalings of the line momenta. With
\begin{equation}
[-(p_1 - q)^2] \sim m_b^2, \quad [m_c^2 - p_1^2] \sim [-p_2^2] \sim [-(p_1 + p_2)^2]  \sim m_c^2\,,
\end{equation}
we can easily deduce that $p_1 \sim p_2 \sim m_c$. The corresponding integral that needs to be expanded according to  $p_1 \sim p_2 \sim m_c \ll m_b$ is then
\begin{align}
&\int_{p_1, p_2}
\frac{1}{[-(p_1-q)^2]} \frac{1}{[m_c^2 - p_1^2]} \frac{1}{[-p_2^2]} \frac{1}{[-(p_1 + p_2)^2] } \nonumber \\
& = \frac{1}{[-m_b^2]} \int_{p_1, p_2} \frac{1}{[m_c^2 - p_1^2]} \frac{1}{[-p_2^2]} \frac{1}{[-(p_1 + p_2)^2] } + \mathcal{O}(m_c^2).
\end{align}

In the second region the inverse propagators scale as
\begin{equation}
[-p_1^2] \sim [m_c^2 - (p_1 + q)^2] \sim [-p_3^2] \sim [-(p_1 + p_3 + q)^2]  \sim m_b^2.
\end{equation}
Without doing any shifts we immediately find that $m_c \ll p_1 \sim p_2 \sim m_b$ so that
\begin{align}
\int_{p_1, p_2}
& \frac{1}{[-p_1^2]} \frac{1}{[m_c^2 - (p_1 + q)^2]} \frac{1}{[-p_2^2]} \frac{1}{[-(p_1 + p_2 + q)^2]}  \nonumber \\
& = \int_{p_1, p_2} \frac{1}{[-p_1^2]} \frac{1}{[- (p_1 + q)^2]} \frac{1}{[-p_2^2]} \frac{1}{[-(p_1 + p_2 + q)^2]} + \mathcal{O}(m_c^2)
\end{align}

Once it is understood how the integral should be expanded in each region, we can proceed to the actual calculation. Instead of doing the corresponding expansions manually, this procedure can be automated by making use of the functionality present in \textsc{FeynCalc} and \textsc{FeynHelpers}.

We need to define a separate \texttt{FCTopology}\footnote{In \textsc{FeynCalc} \texttt{FCTopology} describes the given integral family and contains information about its name, propagators, loop and external momenta as well as kinematic constraints.} object for each region, where the propagators stem from the original integral modulo shifts needed to make the expansion in momenta or masses well defined. Then we introduce a scaling variable $\lambda$ that will be used for expanding the propagators and doing the power counting. Each four-momentum or mass appearing in the integral must be assigned a definite counting by multiplying them with a proper power of $\lambda$.

This can be automated using a routine called \texttt{FCLoopAddScalingParameter}, where the first argument takes the topology and the second one contains a list of scaling rules. All quantities that have not been assigned a scaling rule will be assumed to scale as $\lambda^0$. We also need to declare $\lambda$ as \texttt{FCVariable} to help the code distinguishing between variables and names of momenta inside relevant \textsc{FeynCalc} symbols
\begin{verbatim}
DataType[la,FCVariable]=True;
topoScaledR1=FCLoopAddScalingParameter[topoR1,la,
 {p1->la^1 p1,p2->la^1 p2,q->la^0 q,mb->la^0 mb, mc->la^1 mc}];
topoScaledR2=FCLoopAddScalingParameter[topoR2,la,
 {p1->la^0 p1,p2->la^0 p2,q->la^0 q,mb->la^0 mb, mc->la^1 mc}];
\end{verbatim}

After that we can perform an expansion in $\lambda$ up to the required power with the aid of \texttt{FCLoopGLIExpand}.
While the original integral is written as \texttt{GLI}\footnote{In \textsc{FeynCalc} a \texttt{GLI} denotes a generic loop integral characterized by its name (which must match that of a suitable \texttt{FCTopology}) and the powers of propagators.} symbol, e.g. \texttt{GLI[prop2L,{1,1,1,1}]}, the expansion in $\lambda$ will create a linear combination
of \texttt{GLI}'s with different integer powers of propagators multiplied by scalar products involving loop momenta.

To process this output further, we first apply \texttt{FCLoopFromGLI}, thus converting everything into propagator
representation. Then we handle propagators appearing in the expansion such as $- 2 p_1 \cdot  q + m_b^2$ and alike.
Although \textsc{FeynCalc} can represent pretty arbitrary propagator denominators using \texttt{GFAD} (generic \texttt{FeynAmpDenominator}) containers, most loop-related routines still expect the input to be made of quadratic or 
linear propagators only which are  written as \texttt{SFAD}s (standard \texttt{FeynAmpDenominator}).
The conversion of suitable \texttt{GFAD}s into \texttt{SFAD}s can be automatized using the routine 
\texttt{FromGFAD}, where the user may also add custom rules for cases where the algorithm fails to find a suitable mapping.

After that, the resulting expression can be processed via the standard \textsc{FeynCalc} toolchain for converting a loop amplitude in the propagator representation into a linear combination of \texttt{GLI}s with the corresponding list of \texttt{FCTopology} symbols. To that end we apply \texttt{FCLoop\-Find\-Topologies} (naive topology identification), \texttt{FCLoop\-Basis\-Find\-Completion} and \texttt{FCLoop\-CreateRule\-GLIToGLI} (handling incomplete propagator bases), \texttt{FCLoopFindTopologyMappings} (topology minimization using the algorithm of Ref.~\cite{Pak:2011xt}), \texttt{FCLoopTensorReduce} (tensor reduction) and finally \texttt{FCLoop\-Apply\-Topology\-Mappings} (elimination of explicit loop momenta an introduction of \texttt{GLI}s). The resulting amplitude can be readily IBP-reduced with any suitable software, where for the sake of convenience we choose \textsc{FIRE}~\cite{Smirnov:2014hma,Smirnov:2019qkx,Smirnov:2023yhb}. The \textsc{FeynHelpers} interface to \textsc{FIRE} allows us to carry out the
reduction without leaving the \textsc{Mathematica} notebook, so upon running \texttt{FIREPrepareStartFile}, \texttt{FIRECreateConfigFile}, \texttt{FIRE\-Create\-Integral\-File},
\texttt{FIRECreateLiteRedFiles}, \texttt{FIRECreateStartFile} and finally  \texttt{FIRE\-Run\-Reduction}, we can directly import the reduction tables using \texttt{FIREImportResults}.

Lastly we need to add up contributions from both regions, identify the appearing master integrals and substitute the corresponding analytic expressions. As long as the masters are sufficiently simple to be calculated analytically, this procedure allows us to obtain analytic results for the asymptotic expansion of any suitable integral up to any order. The results can be then checked with the aid of \textsc{FIESTA} and \textsc{pySecDec}.

Sample \textsc{Mathematica} code which illustrates the approach
discussed in this subsection is included in the
supplementary material to this paper.

\subsection{\label{sub::semi}Precise semi-analytic result for physical charm quark masses}

In this subsection we present an efficient algorithm which can be used to compute the master integrals which appear in the $|\Delta B|=1$ theory. For convenience we introduce the linear mass ratio and define 
\begin{eqnarray}
  x&=&\sqrt{z}\,.
\end{eqnarray}

To cover the physically interesting region we apply the ``expand and match''
approach~\cite{Fael:2021kyg,Fael:2022rgm,Fael:2022miw,Fael:2023zqr}.  In the
following we describe the implementation for the problem at hand. The starting
point is the list of master integrals and the corresponding differential
equation in $x$. 
Note that we are only interested in the absorptive part of the three-loop
amplitude. Thus, we eliminate those master integrals from the system which
have no imaginary part. This reduces the system by about 30\%
and leaves us with 342 master integrals.

The basic idea of ``expand and match'' is to construct expansions around
properly chosen values $x_E$ with the help of the differential equations. In
order to determine all expansion coefficients, the expansions have to be
matched to (numerical or analytical) results for the master integrals at a
point $x_M$ which is sufficiently close to $x_E$ such that the series
expansion converges. If $x_E$ is a regular point, i.e., the master integrals
have no threshold for $x=x_E$, it is possible to choose $x_M=x_E$.

The physical amplitude for $B$ meson mixing has cuts through $0,1,2,3$ and $4$
charm quarks and thus we expect thresholds for $x\in \{0,1/4,1/3,1/2,1\}$. In
fact, we observe poles in the differential equations for these values of $x$.
Since our results cover the range $0\le x \lesssim 0.4$ we only have to take
care of first three entries in this list. We observe a good coverage of the
$x$ range in case we add a further expansion point at $x_E=1/10$ such that
we have
\begin{eqnarray}
  x_E \in \{0,1/10,1/4,1/3\}\,.
  \label{eq::xE}
\end{eqnarray}
Depending on the nature of the expansion point a different ansatz has to be
chosen. For our application the most general ansatz for a master integral can
be written as
\begin{eqnarray}
  M_i(x, \epsilon) &=&
  \sum_{j=-2}^{\epsilon_{\mathrm{max}}}
  \sum_{m}^{j+2}  \sum_{n}^{n_{\mathrm{max}}}
  c_{i,j,m,n} \epsilon^j \, (x-x_E)^{n/2} \,
                          \log^m\left(x-x_E\right)\,,
  \label{eq::ansatz}
\end{eqnarray}
where we have used that in the imaginary part of three-loop two-point
functions one has at most $1/\epsilon^2$ poles.  $\epsilon_{\mathrm{max}}$ is
determined by the spurious pole in front of the respective master integral. We
have $\epsilon_{\mathrm{max}}=0$ if they are absent.  Furthermore, we allow
for logarithmic terms with an increase in the power by one unit for each
$\epsilon$ order. These terms are important for the expansions around
$x_E=0,1/4$ and $1/3$.  Finally, $n$ can either take only even values or can
have both even and odd numerical value. The latter is important for the
expansion around $x_M=1/4$ since there are square roots in the expansion
around cuts involving and even number of
massive particles~\cite{Davydychev:1999ic,kmelnikov,Egner:2023kxw}.  For
$x_M=0,1/10,1/3$ only the terms with even $n$ have a non-zero coefficient
$c_{i,j,m,n}$. As upper limit of the summation of $n$ we typically have
$n_{\mathrm{max}}=100$.

For each expansion point we have a dedicated {\tt AMFlow}~\cite{Liu:2022chg} run
for the computation of all master integrals with a precision of 100 digits. We find it convenient to choose
$x_M\in \{1/100,1/10,6/25,29/100\}$.

\begin{figure}[htb]
    \centering
    \includegraphics[scale=0.6]{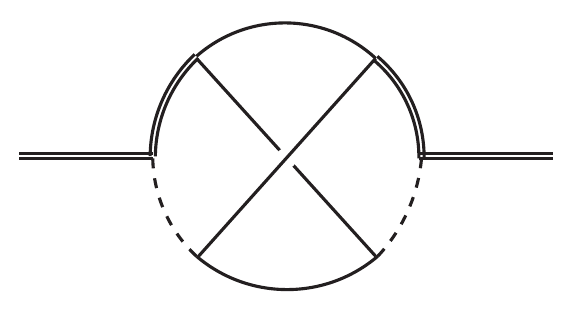}
    \caption{A sample master integral which appears at three-loop with two $Q_{1,2}$ insertions. The dashed, single and double lines denote $m=0$, $m=m_c$ and $m=m_b$ respectively.}
    \label{fig::res_MI_draw}
    %t3l787w0w1w2w3w4w6w8w9w10(1,1,1,1,1,1,1,1,0)
\end{figure}

In Tab.~\ref{tab:precision} we show for three values of $x$
the relative difference of the expansions around the values given in Eq.~(\ref{eq::xE}) for the master integral shown in  Fig.~\ref{fig::res_MI_draw}. We only consider results for the $\epsilon^0$ term of this master integral
which enters the finite term of the physical amplitude.
One observes that for each value there is at least one expansion
which shows an agreement of 20 digits or more. It is expected that the precision gets even better in case $x$ is chosen closer to the expansion point. We note the the precision of the
expansion could be improved by increasing the precision of the boundary integrals or by performing a deeper expansion.

\begin{table}[t]
\centering
\begin{tabular}{@{}  l  l  @{}}
\toprule
Numerical check & Upper bound on relative difference \\
\midrule
At $x=1/20$ & $10^{-43}$ with $x=0$ expansion\\
& $10^{-22}$ with $x=1/10$ expansion \\
\midrule
At $x=1/5$ & $10^{-11}$ with $x=0$ expansion\\
&$10^{-6}$ with $x=1/10$ expansion\\
&$10^{-35}$ with $x=1/4$ expansion\\
\midrule
At $x=3/10$ & $10^{-2}$ with $x=1/4$ expansion\\
& $10^{-30}$ with $x=1/3$ expansion\\
\bottomrule
\end{tabular}
\caption{Relative differences of the series expansions of the integral shown in Fig.~\ref{fig::res_MI_draw} with numerical evaluations in {\tt AMFlow} away from the expansion points.}\label{tab:precision}
\end{table}

As a further check for the precision of our result
we can compare the coefficients of the
$z^0$ and $z^1$ terms of the matching coefficients
$H^{ab}$ and $\widetilde{H}^{ab}_S$ introduced
in Eq.~(\ref{eq::Gam^ab}). Our numerical expressions
agree with the analytic results from Ref.~\cite{Gerlach:2022hoj} at the level of $10^{-40}$ or better.

To illustrate the structure of our results we provide 
explicit results for all 342 master integrals
needed for the computation of the two-point functions
involving current-current operators in an expansion for $x\to 0$. The computer-readable expressions can be downloaded from the website~\cite{progdata}. After the inclusion of penguin operators it might be that there are additional master integrals. However, they can also be computed following the approach described in this Subsection.

%- }}}
%- {{{ Sample result:

\section{\label{sec:results}Sample result}

In the following we briefly 
illustrate the methods described in this paper and give the non-fermionic (i.e. $n_f=0$) NNLO contribution of two $Q_1$ insertions to the matching coefficient $H^{cc}(x)$. The expansion
for $x\to0$ up to $x^{10}$ reads:
\begin{eqnarray}
   H^{cc, (2)}(x) &=& \Big(61.49166 + (334.9196 - 1227.224 \log x + 88.00000 \log^{2} x) \,x^2 - 11.69731 \,x^3
    \nonumber\\&&\mbox{}
    + (-4522.778 + 1030.276 \log x + 376.7253 \log^{2} x) \,x^4 + 199.3985 \,x^5 \nonumber\\&&\mbox{}
    + (-5312.014 + 6117.164 \log x + 1408.251 \log^{2} x) \,x^6 + 500.4536 \,x^7 \nonumber\\&&\mbox{}
    + (-1386.574 + 13388.09 \log x - 773.9628 \log^{2} x) \,x^8 + 1591.938 \,x^9 \nonumber\\&&\mbox{}
    + (-82.72167 + 27995.10 \log x - 1593.620 \log^{2} x) \,x^{10} +{\cal O}(x^{11})\Big)\, C_{1}^2\,.
\end{eqnarray}
Here we have set the renormalisation scale $\mu=m_b$ and the quark masses are renormalised in the pole scheme. Note that the $\Delta B = 2$ evanescent operators were defined as described in Ref.~\cite{Asatrian:2017qaz}. All other $\mathcal{O}(\epsilon)$ contributions of physical operators to evanescent operators that were not previously defined have been set to zero.
The coefficients of $x^0$ and $x^2$ are in agreement with the results obtained in Ref.~\cite{Gerlach:2022hoj}.

%- }}}

%- {{{ Conclusions:

\section{\label{sec:concl}Conclusions}

In this work we address several technical challenges which occur in
the calculation of the parameters necessary to describe the mixing of
neutral $B$ and $D$ mesons. Practical solutions have been presented
for the construction of projectors to scalar functions and for
computing traces with more than 20 $\gamma$ matrices. Furthermore, the
computation of high-rank tensor integrals and the semi-analytic
computation of three-loop master integrals with a dependence on the
ratio of the charm and bottom quark masses has been presented.  Our
algorithms can also be applied to other processes, as, e.g., the
computation of decay rates of $B$ and $D$ mesons.

%- }}}
%- {{{ Acknowledgements:

\section*{Acknowledgements}  

We thank 
Ulrich Nierste, Gurdun Heinrich, Erik Panzer, 
Oliver Schnetz, Kay Sch\"onwald
and Alexander Smirnov for useful discussions. This research was supported by the Deutsche
Forschungsgemeinschaft (DFG, German Research Foundation) under grant 396021762
--- TRR 257 ``Particle Physics Phenomenology after the Higgs Discovery''.

%- }}}

%\newpage
\appendix
\section{Basis elements for spinor structures}
\label{app:basis}

When applying the projectors described in Section \ref{sec:projectors}, we use a subset of the basis elements shown below. As explained previously, we distinguish the following cases:
\begin{enumerate}[label=(\roman*)]
\item Spin line one is longer and spin line two has length $n$. Choose basis with elements up to $n$ $\gamma$ matrices on each spin line as well as the structure with a slashed momentum on spin line one in addition to the $n$ pure $\gamma$ matrices, i.e.~up to and including $B_{4(n+1)-2}$.
\item Spin line two is longer or of the same length and spin line one has length $n$. Choose basis with elements up to $n$ $\gamma$ matrices on each spin line as well as the structure with a slashed momentum on spin line two in addition to the $n$ pure $\gamma$ matrices, i.e.~up to and including $B_{4(n+1)-1}$ but excluding $B_{4(n+1)-2}$.
\item Both spin lines have length eleven and include at least one slashed momentum. This is the longest structure we need to resolve, and we choose a special basis so as not to take unnecessarily long traces. We include all elements up to and including $B_{45}$ ($B_{4(n+1)+1}$ in the notation where $n$ is the number of pure $\gamma$ matrices) but drop $B_{43}$ ($B_{4(n+1)-1}$ in the general case). We do not need $B_{42}$ ($B_{4(n+1)-2}$ in the general case) to project onto, but it is required to make the Gram matrix invertible. This inconvenience arises from the bilinear map $\phi$ defined in Section~\ref{sec:new_phi} not being positive-definite.
\end{enumerate}
The basis elements are:
\begin{align*}
B_1 &= \mathds{1} \otimes \mathds{1}\\
B_2 &= \slashed{e}_q \otimes \mathds{1}\\
B_3 &= \mathds{1} \otimes  \slashed{e}_q\\
B_4 &= \slashed{e}_q \otimes  \slashed{e}_q
\end{align*}
\begin{align*}
B_5 &=  \gamma_{\mu_1}  \otimes  \gamma^{\mu_1}\\
B_6 &=  \gamma_{\mu_1} \slashed{e}_q \otimes  \gamma^{\mu_1}\\
B_7 &=  \gamma_{\mu_1}  \otimes  \gamma^{\mu_1} \slashed{e}_q\\
B_8 &=  \gamma_{\mu_1} \slashed{e}_q \otimes  \gamma^{\mu_1} \slashed{e}_q
\end{align*}
\begin{align*}
B_9 &=  \gamma_{\mu_1} \gamma_{\mu_2}  \otimes  \gamma^{\mu_2} \gamma^{\mu_1}\\
B_{10} &=  \gamma_{\mu_1} \gamma_{\mu_2} \slashed{e}_q \otimes  \gamma^{\mu_2} \gamma^{\mu_1}\\
B_{11} &=  \gamma_{\mu_1} \gamma_{\mu_2} \otimes  \gamma^{\mu_2} \gamma^{\mu_1} \slashed{e}_q\\
B_{12} &=  \gamma_{\mu_1} \gamma_{\mu_2} \slashed{e}_q \otimes  \gamma^{\mu_2} \gamma^{\mu_1} \slashed{e}_q
\end{align*}
\begin{align*}
B_{13} &=  \gamma_{\mu_1} \dots \gamma_{\mu_3}  \otimes  \gamma^{\mu_3} \gamma^{\mu_2} \gamma^{\mu_1}\\
B_{14} &=  \gamma_{\mu_1} \dots \gamma_{\mu_3} \slashed{e}_q \otimes  \gamma^{\mu_3} \gamma^{\mu_2} \gamma^{\mu_1}\\
B_{15} &=  \gamma_{\mu_1} \dots \gamma_{\mu_3}  \otimes  \gamma^{\mu_3} \gamma^{\mu_2} \gamma^{\mu_1} \slashed{e}_q\\
B_{16} &=  \gamma_{\mu_1} \dots \gamma_{\mu_3} \slashed{e}_q \otimes  \gamma^{\mu_3} \gamma^{\mu_2} \gamma^{\mu_1} \slashed{e}_q
\end{align*}
\begin{align*}
B_{17} &=  \gamma_{\mu_1} \dots \gamma_{\mu_4}  \otimes  \gamma^{\mu_4} \dots \gamma^{\mu_1}\\
B_{18} &=  \gamma_{\mu_1} \dots \gamma_{\mu_4} \slashed{e}_q \otimes  \gamma^{\mu_4} \dots \gamma^{\mu_1}\\
B_{19} &=  \gamma_{\mu_1} \dots \gamma_{\mu_4}  \otimes  \gamma^{\mu_4} \dots \gamma^{\mu_1} \slashed{e}_q\\
B_{20} &=  \gamma_{\mu_1} \dots \gamma_{\mu_4} \slashed{e}_q \otimes  \gamma^{\mu_4} \dots \gamma^{\mu_1} \slashed{e}_q
\end{align*}
\begin{align*}
B_{21} &=  \gamma_{\mu_1} \dots \gamma_{\mu_5}  \otimes  \gamma^{\mu_5} \dots \gamma^{\mu_1}\\
B_{22} &=  \gamma_{\mu_1} \dots \gamma_{\mu_5} \slashed{e}_q \otimes  \gamma^{\mu_5} \dots \gamma^{\mu_1}\\
B_{23} &=  \gamma_{\mu_1} \dots \gamma_{\mu_5}  \otimes  \gamma^{\mu_5} \dots \gamma^{\mu_1} \slashed{e}_q\\
B_{24} &=  \gamma_{\mu_1} \dots \gamma_{\mu_5} \slashed{e}_q \otimes  \gamma^{\mu_5} \dots \gamma^{\mu_1} \slashed{e}_q
\end{align*}
\begin{align*}
B_{25} &=  \gamma_{\mu_1} \dots \gamma_{\mu_6}  \otimes  \gamma^{\mu_6} \dots \gamma^{\mu_1}\\
B_{26} &=  \gamma_{\mu_1} \dots \gamma_{\mu_6} \slashed{e}_q \otimes  \gamma^{\mu_6} \dots \gamma^{\mu_1}\\
B_{27} &=  \gamma_{\mu_1} \dots \gamma_{\mu_6}  \otimes  \gamma^{\mu_6} \dots \gamma^{\mu_1} \slashed{e}_q\\
B_{28} &=  \gamma_{\mu_1} \dots \gamma_{\mu_6} \slashed{e}_q \otimes  \gamma^{\mu_6} \dots \gamma^{\mu_1} \slashed{e}_q
\end{align*}
\begin{align*}
B_{29} &=  \gamma_{\mu_1} \dots \gamma_{\mu_7}  \otimes  \gamma^{\mu_7} \dots \gamma^{\mu_1}\\
B_{30} &=  \gamma_{\mu_1} \dots \gamma_{\mu_7} \slashed{e}_q \otimes  \gamma^{\mu_7} \dots \gamma^{\mu_1}\\
B_{31} &=  \gamma_{\mu_1} \dots \gamma_{\mu_7}  \otimes  \gamma^{\mu_7} \dots \gamma^{\mu_1} \slashed{e}_q\\
B_{32} &=  \gamma_{\mu_1} \dots \gamma_{\mu_7} \slashed{e}_q \otimes  \gamma^{\mu_7} \dots \gamma^{\mu_1} \slashed{e}_q
\end{align*}
\begin{align*}
B_{33} &=  \gamma_{\mu_1} \dots \gamma_{\mu_8}  \otimes  \gamma^{\mu_8} \dots \gamma^{\mu_1}\\
B_{34} &=  \gamma_{\mu_1} \dots \gamma_{\mu_8} \slashed{e}_q \otimes  \gamma^{\mu_8} \dots \gamma^{\mu_1}\\
B_{35} &=  \gamma_{\mu_1} \dots \gamma_{\mu_8}  \otimes  \gamma^{\mu_8} \dots \gamma^{\mu_1} \slashed{e}_q\\
B_{36} &=  \gamma_{\mu_1} \dots \gamma_{\mu_8} \slashed{e}_q \otimes  \gamma^{\mu_8} \dots \gamma^{\mu_1} \slashed{e}_q
\end{align*}
\begin{align*}
B_{37} &=  \gamma_{\mu_1} \dots \gamma_{\mu_9}  \otimes  \gamma^{\mu_9} \dots \gamma^{\mu_1}\\
B_{38} &=  \gamma_{\mu_1} \dots \gamma_{\mu_9} \slashed{e}_q \otimes  \gamma^{\mu_9} \dots \gamma^{\mu_1}\\
B_{39} &=  \gamma_{\mu_1} \dots \gamma_{\mu_9}  \otimes  \gamma^{\mu_9} \dots \gamma^{\mu_1} \slashed{e}_q\\
B_{40} &=  \gamma_{\mu_1} \dots \gamma_{\mu_9} \slashed{e}_q \otimes  \gamma^{\mu_9} \dots \gamma^{\mu_1} \slashed{e}_q
\end{align*}
\begin{align*}
B_{41} &=  \gamma_{\mu_1} \dots \gamma_{\mu_{10}}  \otimes  \gamma^{\mu_{10}} \dots \gamma^{\mu_1}\\
B_{42} &=  \gamma_{\mu_1} \dots \gamma_{\mu_{10}} \slashed{e}_q \otimes  \gamma^{\mu_{10}} \dots \gamma^{\mu_1}\\
B_{43} &=  \gamma_{\mu_1} \dots \gamma_{\mu_{10}}  \otimes  \gamma^{\mu_{10}} \dots \gamma^{\mu_1} \slashed{e}_q\\
B_{44} &= \gamma_{\mu_1} \dots \gamma_{\mu_{10}} \slashed{e}_q \otimes  \gamma^{\mu_{10}} \dots \gamma^{\mu_1} \slashed{e}_q
\end{align*}
\begin{align}
    B_{45} &=  \gamma_{\mu_1} \dots \gamma_{\mu_{11}}  \otimes  \gamma^{\mu_{11}} \dots \gamma^{\mu_1}
\end{align}
Here, we defined
\begin{equation}
\slashed{e}_q \equiv \frac{\slashed{q}}{\sqrt{q^2}},
\end{equation}

%%%\bibliographystyle{JHEP} 
%%%\footnotesize
%%%\bibliography{BIB}

%- {{{ Effective operators

\section{\label{app::ops}Effective operators}

For completeness we provide in this Appendix all relevant operator
for the $|\Delta B|=1$ and $|\Delta B|=2$ theories.

On the $|\Delta B|=1$ side we work with the operator basis
first introduced in Ref.~\cite{Chetyrkin:1997gb}
\begin{eqnarray}
	\mathcal{H}_{\textrm{eff}}^{|\Delta B|=1} 
	&=&   \frac{4G_F}{\sqrt{2}}  \left[
	-\, \lambda^s_t \Big( \sum_{i=1}^6 C_i Q_i + C_8 Q_8 \Big) 
	- \lambda^s_u \sum_{i=1}^2 C_i (Q_i - Q_i^u) \right. \nonumber\\
	&& \phantom{\frac{4G_F}{\sqrt{2}} \Big[}
	\left.
	+\, V_{us}^\ast V_{cb} \, \sum_{i=1}^2 C_i Q_i^{cu} 
	+ V_{cs}^\ast V_{ub} \, \sum_{i=1}^2 C_i Q_i^{uc} 
	\right]
	+ \mbox{h.c.}\,,
	\label{eq::HamDB1}
\end{eqnarray}
with
\begin{eqnarray}
  \lambda^s_a = V_{as}^\ast V_{ab}\,, 
\end{eqnarray}
where $a=u,c,t$ and $\lambda_t=-\lambda_c-\lambda_u$.  $G_F$ stands for the
Fermi constant. The physical operators read
\begin{eqnarray}
	Q_1   &=& \bar{s}_L \gamma_{\mu} T^a c_L\;\bar{c}_L     \gamma^{\mu} T^a b_L\,,\nonumber\\
	Q_2   &=& \bar{s}_L \gamma_{\mu}     c_L\;\bar{c}_L     \gamma^{\mu}     b_L\,,\nonumber\\
	Q_3   &=& \bar{s}_L \gamma_{\mu}     b_L \sum_q \bar{q}\gamma^{\mu}     q\,,\nonumber\\
	Q_4   &=& \bar{s}_L \gamma_{\mu} T^a b_L \sum_q \bar{q}\gamma^{\mu} T^a q\,,\nonumber\\
	Q_5   &=& \bar{s}_L \gamma_{\mu_1}\gamma_{\mu_2}\gamma_{\mu_3} b_L
                  \sum_q \bar{q} \gamma^{\mu_1} \gamma^{\mu_2}\gamma^{\mu_3}     q\,,\nonumber\\
	Q_6   &=& \bar{s}_L \gamma_{\mu_1}\gamma_{\mu_2}\gamma_{\mu_3} T^a b_L
                  \sum_q \bar{q} \gamma^{\mu_1}\gamma^{\mu_2}\gamma^{\mu_3} T^a q\,,\nonumber\\
	Q_8   &=& \frac{g_s}{16\pi^2} m_b \, \bar{s}_L \sigma^{\mu \nu} T^a b_R \, G_{\mu\nu}^a\,, \nonumber\\
	Q^u_1 &=& \bar{s}_L \gamma_{\mu} T^a u_L\;\bar{u}_L     \gamma^{\mu} T^a b_L\,,\nonumber \\
	Q^u_2 &=& \bar{s}_L \gamma_{\mu}     u_L\;\bar{u}_L     \gamma^{\mu}     b_L\,,\nonumber\\
	Q^{cu}_1 &=& \bar{s}_L \gamma_{\mu} T^a u_L\;\bar{c}_L     \gamma^{\mu} T^a b_L\,,\nonumber \\
	Q^{cu}_2 &=& \bar{s}_L \gamma_{\mu}     u_L\;\bar{c}_L     \gamma^{\mu}     b_L\,,\nonumber\\
	Q^{uc}_1 &=& \bar{s}_L \gamma_{\mu} T^a c_L\;\bar{u}_L     \gamma^{\mu} T^a b_L\,,\nonumber \\
	Q^{uc}_2 &=& \bar{s}_L \gamma_{\mu}     c_L\;\bar{u}_L     \gamma^{\mu}     b_L\,.
	\label{operators}
\end{eqnarray}
Here we introduced the left-chiral projector $P_L=(1-\gamma_5)/2$ so that
$q_L=P_Lq$. We have current-current operators $Q_{i}, Q_{i}^{u}, Q_{i}^{uc}$
and $Q_{i}^{cu}$ ($i=1,2$), four-quark penguin operators $Q_3,\ldots,Q_6$ and
the chromomagnetic penguin operator $Q_8$ featuring the gluon field strength
$G^a_{\mu \nu}$, the strong coupling $g_s$ and
$\sigma^{\mu \nu} = i[\gamma^\mu,\gamma^\nu]/2$.  In the definitions of
four-quark penguin operators the sum over $q$ is understood to run over all
five flavors $u,d,s,c$ and $b$.

The LO evanescent operators required at NLO accuracy and beyond read
\begin{eqnarray}
	E_1[Q_1] &=& 
	\bar{s}_L \gamma^{\mu_1} \gamma^{\mu_2} \gamma^{\mu_3}
	T^a c \;
	\bar{c} \gamma_{\mu_1} \gamma_{\mu_2} \gamma_{\mu_3}  T^a
	b_L - 16 Q_1\,,\nonumber\\ 
	E_1[Q_2] &=& 
	\bar{s}_L \gamma^{\mu_1} \gamma^{\mu_2} \gamma^{\mu_3}
	c_i\;\bar{c}_j \gamma_{\mu_1} \gamma_{\mu_2}
	\gamma_{\mu_3} b_L - 16 Q_2\,,\nonumber\\ 
	E_1[Q_5] &=& 
	\bar{s}_L \gamma^{\mu_1} \gamma^{\mu_2} \gamma^{\mu_3}
	\gamma^{\mu_4} \gamma^{\mu_5} b_L \sum_q \bar{q}
	\gamma_{\mu_1} \gamma_{\mu_2} \gamma_{\mu_3} \gamma_{\mu_4}
	\gamma_{\mu_5} q_j \, - \, 20 Q_5 \, +\,  64 Q_3\,,\nonumber\\ 
	E_1[Q_6] &=& 
	\bar{s}_L \gamma^{\mu_1} \gamma^{\mu_2} \gamma^{\mu_3}
	\gamma^{\mu_4} \gamma^{\mu_5} T^a b_L \sum_q
	\bar{q} \gamma_{\mu_1} \gamma_{\mu_2} \gamma_{\mu_3}
	\gamma_{\mu_4} \gamma_{\mu_5} T^a q \,-\, 20 Q_6 \,+\, 64 Q_4 
	\, .
	\label{evan_operators_lo}
\end{eqnarray}
Similar evanescent operators are needed for $Q_{i}^{u}$, $Q_{i}^{uc}$ and
$Q_{i}^{cu}$ with $i=1,2$; they are obtained by replacing one or both $c$
fields with $u$ fields in Eq.~(\ref{evan_operators_lo}).

At NNLO one has to introduce the so-called
NLO evanescent operators which are given by~\cite{Chetyrkin:2017mwp}
\begin{eqnarray}
	E_2[Q_1] &=& 
	\bar{s}_L \gamma^{\mu_1} \gamma^{\mu_2} \gamma^{\mu_3} \gamma^{\mu_4} \gamma^{\mu_5}
	T^a c \;
	\bar{c} \gamma_{\mu_1} \gamma_{\mu_2} \gamma_{\mu_3} \gamma^{\mu_4} \gamma^{\mu_5}  T^a
	b_L - 20 E_1 [Q_2] - 256 Q_1\,,\nonumber\\ 
	E_2[Q_2] &=& 
	\bar{s}_L \gamma^{\mu_1} \gamma^{\mu_2} \gamma^{\mu_3} \gamma^{\mu_4} \gamma^{\mu_5}
	c_i\;\bar{c}_j \gamma_{\mu_1} \gamma_{\mu_2}
	\gamma_{\mu_3} \gamma^{\mu_4} \gamma^{\mu_5} b_L - 20 E_1 [Q_2] - 256 Q_2\,,\nonumber\\ 
	E_2[Q_5] &=& 
	\bar{s}_L \gamma^{\mu_1} \gamma^{\mu_2} \gamma^{\mu_3}
	\gamma^{\mu_4} \gamma^{\mu_5} \gamma^{\mu_6} \gamma^{\mu_7} b_L \sum_q \bar{q}
	\gamma_{\mu_1} \gamma_{\mu_2} \gamma_{\mu_3} \gamma_{\mu_4}
	\gamma_{\mu_5} \gamma^{\mu_6} \gamma^{\mu_7} q_j \, - \, 336 Q_5 \, +\,  1280 Q_3\,,\nonumber\\ 
	E_2[Q_6] &=& 
	\bar{s}_L \gamma^{\mu_1} \gamma^{\mu_2} \gamma^{\mu_3}
	\gamma^{\mu_4} \gamma^{\mu_5} \gamma^{\mu_6} \gamma^{\mu_7} T^a b_L \sum_q
	\bar{q} \gamma_{\mu_1} \gamma_{\mu_2} \gamma_{\mu_3}
	\gamma_{\mu_4} \gamma_{\mu_5} \gamma^{\mu_6} \gamma^{\mu_7} T^a q \nonumber \\
	&&\mbox{} - 336 Q_6 \,+\, 1280 Q_4 
	\, .
	\label{evan_operators_nlo}
\end{eqnarray}

In the $|\Delta B| =2$ theory there two physical operators which we choose as
\begin{eqnarray}
	Q &=& \bar{s}_i \gamma^\mu \,(1-\gamma^5)\, b_i \; \bar{s}_j \gamma_\mu
	\,(1-\gamma^5)\, b_j\,, \nonumber\\
	\widetilde{Q}_S &=& \bar{s}_i \,(1+\gamma^5)\, b_j\; \bar{s}_j \,(1+\gamma^5)\,
	b_i\,, 
                            \label{eq::opDB2}
\end{eqnarray}
where $i,j$ are the colour indices attached to quark fields.

%- }}}

\bibliographystyle{jhep} 
\bibliography{inspire.bib}

%\begin{thebibliography}{99}
%
%\input{bmix_tech_ref}
%
%\end{thebibliography}

\end{document}